\documentclass[reprint,showkeys,showpacs,superscriptaddress,floatfix]{revtex4-2}
\usepackage{amsmath}
\usepackage{amssymb}
\usepackage{graphicx}
\usepackage{hyperref}
\hypersetup{
	colorlinks=true,
	citecolor=blue,
	linkcolor=red,
	filecolor=blue,      
	urlcolor=blue,
}

\begin{document}

\title{\texorpdfstring{Griffiths-like phase, spin-phonon coupling, and exchange-bias in the disordered double perovskite GdSrCoMnO$_{6}$}{Griffiths-like phase, spin-phonon coupling, and exchange-bias in the disordered double perovskite GdSrCoMnO6}}

\author{Gyanti Prakash Moharana}
\email{gyantiprakashm@gmail.com}
\affiliation{Institute of Physics, Sachivalaya Marg, Bhubaneswar 751005, India}
\affiliation{Homi Bhabha National Institute, Anushakti Nagar, Mumbai 400085, India}

\author{Diptikanta Swain}
\affiliation{Institute of Chemical Technology--IndianOil Odisha Campus, Bhubaneswar 751013, India}

\author{Hanuma Kumar Dara}
\affiliation{Institute of Physics, Sachivalaya Marg, Bhubaneswar 751005, India}
\affiliation{Homi Bhabha National Institute, Anushakti Nagar, Mumbai 400085, India}

\author{Debendra Prasad Panda}
\affiliation{School of Advanced Materials, and Chemistry and Physics of Materials Unit, Jawaharlal Nehru Centre for Advanced Scientific Research, Jakkur, Bengaluru 560064, India}

\author{S. N Sarangi}
\affiliation{Institute of Physics, Sachivalaya Marg, Bhubaneswar 751005, India}
\affiliation{Homi Bhabha National Institute, Anushakti Nagar, Mumbai 400085, India}

\date{\today}

\begin{abstract}
We report structural, magnetic, and Raman studies of the disordered double perovskite GdSrCoMnO$_{6}$~(GSCM). DC magnetization shows a ferromagnetic transition at $T_{C} \approx 153$~K. The inverse susceptibility exhibits a downturn above $T_{C}$ and is consistent with a Griffiths-like regime extending up to $T_{G} \approx 172$~K. Raman measurements show a deviation of the phonon frequency from the anharmonic background near the magnetic-ordering region, consistent with spin-phonon coupling. AC susceptibility indicates slow magnetic dynamics below the freezing temperature $T_{f} \approx 30$~K. These results point to magnetic inhomogeneity generated by the random distribution of mixed-valence Co and Mn ions and by the resulting competition between ferromagnetic and antiferromagnetic interactions. In the low-temperature regime, an exchange-bias effect is observed up to 50~K, with an exchange-bias magnitude $|H_{EB}| = 379$~Oe at 5~K. Structural disorder therefore plays an important role in the magnetic correlations, spin dynamics, and spin-lattice response of GSCM.
\end{abstract}


\maketitle
\section{Introduction}

Oxide perovskites and double perovskites remain a useful platform for studying coupled spin, charge, orbital, and lattice degrees of freedom~\cite{tokura2006critical}. Double perovskites, in particular, exhibit a broad range of magnetic and functional responses, including multiferroicity, spin-phonon coupling, metal-insulator transitions, colossal magnetoresistance, and exchange bias~\cite{mahato2010colossal,nair2011griffiths,macedo2013spin,zhao2014near,su2015magnetism,neenu2015colossal,meng2016microscopic,wang2015effect}. A central issue in these materials is crystallographic disorder, especially antisite disorder. Such disorder modifies the balance between competing ferromagnetic (FM) and antiferromagnetic (AFM) exchange paths and can generate magnetically inhomogeneous states~\cite{nair2014magnetization,murthy2015giant,singh2016influence}.

In rare-earth double perovskites, changes at the $A$ site provide an effective way to tune the lattice geometry and the magnetic exchange network. In compounds of the form R$_{2}$(Co/Mn)O$_{6}$, ionic substitution changes bond lengths and bond angles and therefore affects the dominant exchange interactions and the magnetic ordering temperature~\cite{nasir2019role}. Such lattice changes can also favor Griffiths-like behavior, in which short-range FM clusters develop in a globally paramagnetic matrix and produce a downturn in $\chi^{-1}(T)$ above the bulk ordering temperature~\cite{das2020non}. Related Griffiths-phase phenomenology has been reported in several correlated oxides and disordered magnetic systems, which highlights the role of quenched disorder and phase competition in shaping the magnetic response~\cite{salamon2002colossal,tong2008griffiths,jiang2008griffiths,deisenhofer2005observation,lu2009ru,qian2008enhancement,yang2007observation,shimada2006semiconducting,liu2014griffiths,singh2018structural,salamon2003griffiths,neto1998non,das2020non}.

Co-, Mn-, and Ni-based double perovskites are especially relevant in this context. In these systems, antisite disorder and mixed valence can promote competing FM and AFM interactions, slow magnetic dynamics, interfacial frustration, and exchange-bias effects. Exchange bias is usually associated with coupling between magnetically distinct regions, including FM/AFM, FM/ferrimagnetic, FM/spin-glass, and AFM/spin-glass interfaces~\cite{yanez2011multiferroic,wang2015effect,chikara2016electric}. Previous studies on rare-earth double perovskites have reported Griffiths-like behavior, spin-glass dynamics, magnetization reversal, multiferroicity, and magnetodielectric effects~\cite{bhatti2019magnetic,liu2014griffiths,choudhury2012near,banerjee2018magnetization,madhogaria2019evidence,blasco2015evidence,moon2018anisotropic,sahoo2019exchange,silva2022griffiths,silva2023magnetoelastic}. Sr substitution in related compounds has also been shown to increase antisite disorder, strengthen competing interactions, and enhance exchange-bias responses~\cite{sahoo2019exchange,silva2022griffiths,silva2023magnetoelastic}.

Gd-based double perovskites are particularly interesting because the transition-metal $3d$ moments coexist with rare-earth $4f$ moments~\cite{murthy2015giant,moon2017giant}. In Gd$_{2}$CoMnO$_{6}$, the ferromagnetic transition is governed mainly by the transition-metal exchange network, whereas the low-temperature response is influenced by the Gd sublattice and by disorder-driven competing interactions~\cite{murthy2015giant,moon2017giant}. Replacing Gd$^{3+}$ with Sr$^{2+}$ is expected to modify the lattice as well as the transition-metal valence balance. It should therefore affect the disorder level, the exchange paths, and the tendency toward magnetically inhomogeneous states. GdSrCoMnO$_{6}$ is thus a suitable system for examining how structural disorder, mixed valence, and lattice dynamics are coupled in a disordered double perovskite.

In this work, we investigate the structural, magnetic, and Raman properties of GdSrCoMnO$_{6}$ under ambient conditions. The compound adopts a disordered orthorhombic structure and shows ferromagnetic ordering near $T_{C}\approx153$~K, which is higher than that of the parent compound Gd$_{2}$CoMnO$_{6}$. The inverse susceptibility shows a downturn above $T_{C}$ that is consistent with a Griffiths-like regime extending up to $T_{G}\approx172$~K. Raman measurements reveal a deviation from the anharmonic phonon background near the magnetic-ordering region, consistent with spin-phonon coupling. AC susceptibility indicates slow low-temperature glassy dynamics below $T_{f}\approx30$~K. In the same temperature range, the material exhibits exchange bias associated with competing FM and AFM interactions. These results provide a framework for understanding the interplay of disorder, magnetic inhomogeneity, and spin-lattice coupling in Sr-substituted Gd-based double perovskites.

\section{Experimental Methods}

The single-phase polycrystalline GSCM sample was synthesized by a conventional solid-state reaction method. High-purity powders of Gd$_{2}$O$_{3}$ (99.9$\%$), Co$_{3}$O$_{4}$ (99.9$\%$), SrCO$_{3}$, and MnO$_{2}$ (99.9$\%$) were used as starting materials~\cite{das2020non}. Gd$_{2}$O$_{3}$ was preheated at 950$^{\circ}$C for 24~h to remove moisture. The powders were mixed in stoichiometric proportion and heated at 1200$^{\circ}$C in an oxygen-rich environment with intermediate grindings. The product was then reground, pelletized, and sintered at 1350$^{\circ}$C for 24~h under ambient conditions.

The phase purity and average crystal structure were examined by high-resolution x-ray powder diffraction using CuK$_{\alpha}$ radiation (Rigaku TTRAX II, $\lambda=1.5406$~\AA) at room temperature. The diffraction peaks were indexed with an orthorhombic $Pnma$ structure, and no impurity phase was detected within the sensitivity of the measurement. Selected-area electron diffraction (SAED) and high-resolution transmission electron microscopy (HRTEM) were used to examine the local structural and microstructural features. These measurements were performed using a Tecnai G2 F20 S-Twin transmission electron microscope operated at 200~kV. Magnetic properties were measured by DC and AC magnetization using a superconducting quantum interference device vibrating sample magnetometer (Quantum Design). Temperature-dependent DC magnetization was recorded from 5 to 300~K under different applied fields. Field-cooled (FC) and zero-field-cooled (ZFC) hysteresis loops were measured over a field range of $\pm$50~kOe.

\section{Results and Discussion}
\subsection{Structural and Microstructural Study}
\begin{figure}[!t]
	\centering
	\includegraphics[width=\columnwidth]{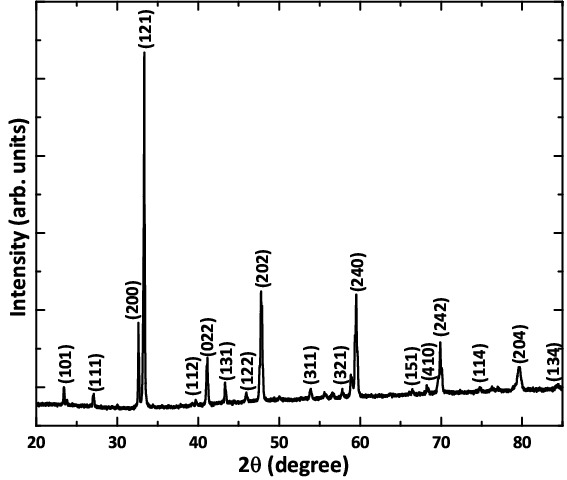}
	\caption{(a) Powder X-ray diffraction pattern recorded at 300~K, and (b) schematic crystal structure of GSCM generated using VESTA software.}
	\label{xrd}
\end{figure}
\begin{figure}[!t]
	\centering
	\includegraphics[width=\columnwidth]{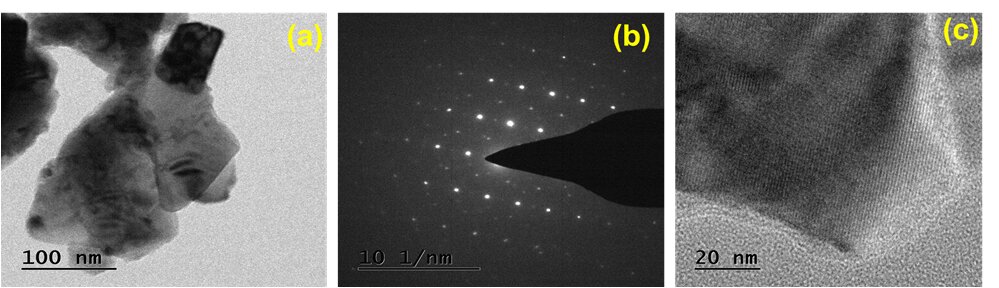}
	\caption{(a) Typical transmission electron microscope (TEM) image, (b) selected-area electron diffraction (SAED) pattern, and (c) corresponding high-resolution transmission electron microscope (HRTEM) image showing lattice fringes of the GSCM sample.}
	\label{tem}
\end{figure}

The crystalline quality and phase purity of the sample were investigated by X-ray diffraction, as shown in Fig.~\ref{xrd}. No trace of any secondary phase was detected within the sensitivity limit of the instrument, confirming the high quality and single-phase nature of the GSCM compound. Fig.~\ref{tem}(a) shows a TEM image of the sample, from which an average particle size of $\sim$250~nm is estimated. The selected-area electron diffraction pattern shown in Fig.~\ref{tem}(b) indicates that the individual particle examined is single crystalline. In addition, clear lattice fringes are observed in the HRTEM image shown in Fig.~\ref{tem}(c).

\subsection{DC Magnetization Study}
\begin{figure*}[!t]
	\centering
	\includegraphics[width=0.82\textwidth]{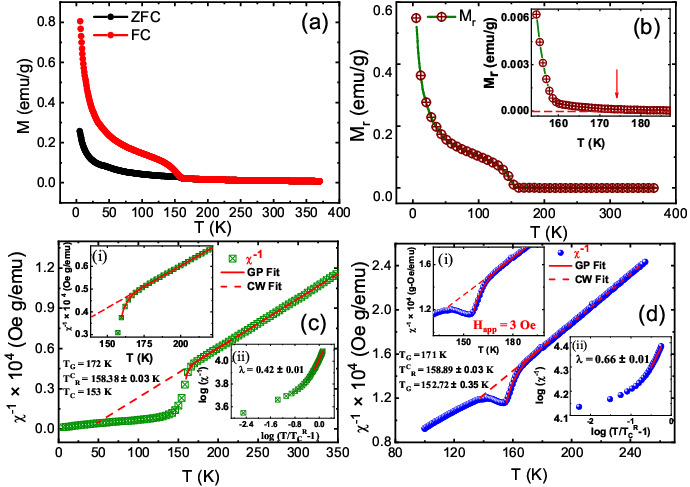}
	\caption{Temperature variation of (a) magnetization in the ZFC and FC processes under an applied field of 100~Oe, (b) thermomagnetic irreversibility $(M_{FC}-M_{ZFC})$; the inset shows its disappearance above 175~K. (c) The solid line represents the best fit of Eq.~\ref{eq:GP} to the $\chi^{-1}(T)$ data obtained from the DC magnetization study, with parameters $\mathrm{T}_{C}^{R}$ = 158.89~K and $\lambda$ = 0.42. A log-log plot in the inset shows the power-law behavior in $\chi^{-1}(T)$. (d) The solid line represents the best fit of Eq.~\ref{eq:GP} to the $\chi^{-1}(T)$ data obtained from the zero-field AC susceptibility study, with parameters $\mathrm{T}_{C}^{R}$ = 158.38~K and $\lambda$ = 0.66. A log-log plot in the inset demonstrates the power-law behavior in $\chi^{-1}(T)$.}
	\label{mag}
\end{figure*}

Fig.~\ref{mag}(a) shows the field-cooled (FC) and zero-field-cooled (ZFC) magnetization measured at 100~Oe. The magnetization rises sharply near 153~K, which marks the onset of ferromagnetic order. The obtained $T_{C}$ is higher than that reported for the parent compound Gd$_{2}$CoMnO$_{6}$ ($T_{C}=123$~K)~\cite{murthy2015giant,moon2017giant}. This enhancement may be associated with mixed-valence Co and Mn states, which can modify the exchange paths and favor ferromagnetic alignment. Sr substitution may also change the local bonding geometry and the effective hopping amplitudes. Both effects can stabilize ferromagnetic order and increase $T_{C}$. The increase in magnetization at low temperature [Fig.~\ref{mag}(a)] is attributed to the weakly interacting Gd$^{3+}$ moments, which carry a total spin $S=7/2$. Figure~\ref{mag}(b) shows the thermomagnetic irreversibility, defined here as $(M_{FC}-M_{ZFC})$. This quantity vanishes near 172~K. The FC/ZFC bifurcation up to this temperature is consistent with short-range ferromagnetic clusters embedded in a paramagnetic matrix~\cite{das2020non,wu2003glassy}.

To further verify this behavior, the high-temperature inverse-susceptibility data were analyzed using the Curie--Weiss law ($\chi=C/(T-\theta)$) for the DC magnetization data of GSCM recorded at 100~Oe. The obtained parameters are the Curie constant, $C = 0.07$~emu.K$/$g.Oe, and the Curie--Weiss temperature, $\theta = 25$~K. Note that $\theta$ (= 25~K) is much smaller than $T_{C}$ (= 153~K). This behavior may reflect the highly disordered nature of the system, arising from the random distribution of ions~\cite{bull2016magnetic}. However, a clear deviation from Curie--Weiss behavior is observed below 172~K, which is identified as the Griffiths temperature, $T_{G}$.

The random arrangement of multivalent magnetic ions in disordered double perovskites induces phase heterogeneity and quenched disorder, manifesting as valence-state fluctuations, as expressed in Eq.~\ref{eq:Oxi}:
\begin{equation}
	Mn^{4+}+Co^{2+}\leftrightharpoons Mn^{3+}+Co^{3+}
	\label{eq:Oxi}
\end{equation}
This results in antisite disorder (ASD), which can be phenomenologically estimated using Eq.~\ref{eq:Qunat}, a form commonly employed for antisite-disordered double perovskites~\cite{nair2014magnetization,sahoo2019exchange}:
\begin{equation}
	\begin{split}
	  M_{S}(ASD, x)= (1-2ASD)(M_{Co}+M_{Mn}) \\
	  ~+~x(2ASD-1)
	\end{split}
	\label{eq:Qunat}
\end{equation}
Using the above expression, the ASD level is phenomenologically estimated to be of order 25$\%$, consistent with the strong disorder scenario commonly invoked for Co/Mn double perovskites~\cite{nair2014magnetization,bull2016magnetic,murthy2015giant}. This value should be viewed as indicative rather than definitive, since antisite disorder is not measured directly here. Enhanced disorder is expected to generate multiple magnetic regions and to increase frustration. The possible exchange paths include Co$^{2+}$-O$^{2-}$-Mn$^{4+}$~(FM), Co$^{3+}$-O$^{2-}$-Mn$^{3+}$~(FM), Co$^{2+}$-O$^{2-}$-Co$^{2+}$~(AFM), Mn$^{4+}$-O$^{2-}$-Mn$^{4+}$~(AFM), and Mn$^{3+}$-O$^{2-}$-Mn$^{3+}$~(AFM). Such competition provides a natural basis for a cluster-glass-like state and for exchange bias. These issues are discussed further below.

Because mixed valence, antisite disorder, possible multiple spin states of Co, and the additional contribution from the Gd$^{3+}$ sublattice complicate a unique microscopic determination of the paramagnetic moment in GSCM, we do not interpret the magnetization data in terms of a single definitive ionic-moment model. Nevertheless, to provide a qualitative reference for the local-moment scale expected in a representative mixed-valence scenario, we consider an illustrative composition containing Co$^{2+}$/Co$^{3+}$ and Mn$^{3+}$/Mn$^{4+}$ fractions of 0.25/0.75 and 0.25/0.75, respectively. Within this illustrative picture, the corresponding moment scale may be written as Eq.~\ref{eq:mag_scenario}:
\begin{equation}
\begin{split}
\mathrm{\mu}_{scen}^{2}=[(\mu_{Gd^{3+}})^{2}~+~(0.75\mu_{Co^{3+}}+0.25\mu_{Co^{2+}})^{2} \\ 
~+~(0.25\mu_{Mn^{3+}}+0.75\mu_{Mn^{4+}})^{2}]~\mu_{B}~~~~~~~
\label{eq:mag_scenario}
\end{split}	
\end{equation}
where $\mu_{Gd}=g_{J}\sqrt{J(J+1)}\,\mu_{B}$ and $\mu_{Co/Mn}=g_{s}\sqrt{S(S+1)}\,\mu_{B}$. The values listed in Table.~\ref{tab:mag} should therefore be regarded only as an illustrative mixed-valence local-moment scenario and not as a direct quantitative counterpart of the effective moments extracted from the Curie--Weiss analysis.

\begin{figure*}[!t]
	\centering
	\includegraphics[width=0.82\textwidth]{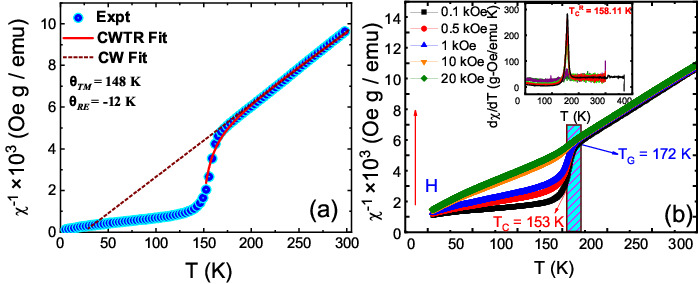}
	\caption{(a) Best fit to 1/$\chi(T)$ using the modified Curie--Weiss formula [Eq.~\ref{eq:MCW}], and (b) $\chi^{-1}(T)$ curves recorded at different magnetic fields (0.01 to 20~kOe). The inset shows the $T_{C}^{R}$ value.}
	\label{mcw_fit}
\end{figure*}

Because the simple Curie--Weiss fit reflects the combined response of the transition-metal and rare-earth sublattices, it does not by itself isolate the magnetic correlations associated with the transition-metal network. We therefore also analyze the data using a modified Curie--Weiss expression that separates the transition-metal and rare-earth contributions. To examine the inverse-susceptibility downturn more appropriately, we adopt a modified Curie--Weiss expression,
\begin{equation}
	\chi(T)=\chi_{TM}+\chi_{R}=\frac{C_{TM}}{T-\theta_{TM}}+\frac{C_{R}}{T-\theta_{R}}
	\label{eq:MCW}
\end{equation}
where $\chi_{TM}$ and $\chi_{R}$ denote the transition-metal and rare-earth contributions, respectively, and $C_{TM}$, $\theta_{TM}$, $C_{R}$, and $\theta_{R}$ are the corresponding Curie constants and Curie--Weiss temperatures. The best fit of Eq.~\ref{eq:MCW} is shown in Fig.~\ref{mcw_fit}(a). The fit yields $\theta_{TM}=148$~K, indicating that the dominant transition-metal magnetic correlations are closely associated with the ferromagnetic transition. The effective paramagnetic moment obtained from the transition-metal Curie constant is $\mu_{eff}=5.59$~$\mu_{B}$/f.u., which is broadly consistent with a simplified Co$^{2+}$/Mn$^{4+}$-dominated estimate [$(\mu_{eff})_{theo}=5.47$~$\mu_{B}$/f.u.], as reported for related systems~\cite{murthy2015giant}. Likewise, the fit gives $C_{Gd}=0.492$ and $\theta_{Gd}=-11.78$~K for the rare-earth-related contribution, with the negative $\theta_{Gd}$ indicating antiferromagnetic alignment relative to the transition-metal sublattice. These parameters should therefore be regarded as effective quantities within the adopted two-sublattice description of the coupled magnetic response.

\begin{table*}[!t]
	\centering
	\caption{Illustrative ionic moments for one representative mixed-valence/spin-state scenario considered for GSCM. These values are included only as a qualitative aid and are not directly compared with the effective moments extracted from the modified Curie--Weiss analysis.}
	\label{tab:mag}
	\begin{tabular}{c c c c c c}
		\hline\hline
		Gd$^{3+}$ ($\mu_{B}$) & (0.25)~Mn$^{3+}$ ($\mu_{B}$) & (0.75)~Mn$^{4+}$ ($\mu_{B}$)  & (0.75)~Co$^{3+}$ ($\mu_{B}$) & (0.25)~Co$^{2+}$ ($\mu_{B}$) & Representative scenario \\
		\hline
		$\mathrm{J}$ = $\frac{7}{2}$~(HS) & $\mathrm{S}$ = 2~(HS) & $\mathrm{S}$ = $\frac{3}{2}$~(HS) & $\mathrm{S}$ = 1~(HS) & $\mathrm{S}$ = $\frac{3}{2}$~(HS) & --- \\
		\hline
		7.93 & 4.90 & 3.87 & 2.82 & 3.87 & 9.21 \\
		\hline\hline
	\end{tabular}
\end{table*}

\subsection{Griffiths Phase Analysis}
The pronounced downturn in $\chi^{-1}(T)$ above $T_{C}$, as shown in Fig.~\ref{mag}(c), indicates the development of short-range ferromagnetic correlations in the paramagnetic background and is consistent with a Griffiths-like regime in GSCM. To analyze this behavior, the inverse susceptibility was examined using the standard Griffiths-phase power-law expression $\chi^{-1} \propto (T/T_{C}^{R} -1)^{1-\lambda}$~\cite{salamon2002colossal,deisenhofer2005observation}, where $0 < \lambda < 1$ characterizes the deviation from ideal Curie--Weiss behavior.
\begin{equation}
	\chi^{-1} \propto\left( \frac{T}{\mathrm{T}_{C}^{R}} -1\right)^{1-\lambda}
	\label{eq:GP}
\end{equation}
The analysis of the DC susceptibility yields $T_{C}^{R} = 158.89 \pm 0.09$~K and $\lambda = 0.42 \pm 0.01$, while the corresponding analysis of the zero-field AC susceptibility gives $T_{C}^{R} = 158.38 \pm 0.03$~K and $\lambda = 0.66 \pm 0.01$. In contrast, the higher-temperature paramagnetic region gives $\lambda_{PM} = -0.05 \pm 0.15$, indicating that the anomalous downturn is not sustained in the normal paramagnetic state. Thus, the temperature interval between $T_{C} \approx 153$~K and $T_{G} \approx 172$~K may be identified as a Griffiths-like regime associated with magnetic inhomogeneity and short-range clustered correlations. The progressive suppression of the downturn in $\chi^{-1}(T)$ with increasing magnetic field further supports this interpretation~\cite{deisenhofer2005observation,liu2014griffiths}.

To probe the low-temperature dynamics associated with this inhomogeneous state, the AC susceptibility analysis is discussed in the following section.

\subsection{AC magnetization study}
AC susceptibility ($\chi^{'}$- real part and $\chi^{"}$- imaginary part) measurements on the GSCM sample were carried out over the temperature range 5~K--200~K using different frequencies (399~Hz to 799~Hz) at $H_{AC}$ = 3~Oe, as shown in Fig.~\ref{ac-chi}. To quantify the frequency-dependent freezing behavior, the Mydosh parameter ($\Phi_{f}$) is introduced in Eq.~\ref{eq:Mydosh}~\cite{mydosh1993spin,nair2011griffiths,das2020non}:

\begin{figure*}[!t]
	\centering
	\includegraphics[width=0.82\textwidth]{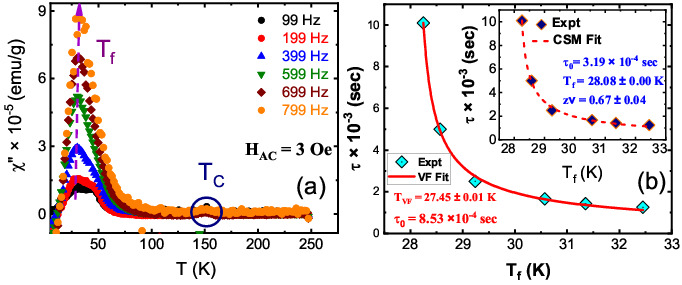}
	\caption{(a) $\chi^{"}(T)$ of the AC susceptibility as a function of frequency, and (b) Vogel-Fulcher plot for the investigation of the glassy system. The inset shows the critical slowing down model fit to the experimental data.}
	\label{ac-chi}
\end{figure*}

\begin{equation}
	\Phi_{f} = \frac{\Delta T_{f}}{T_{f}\Delta log_{10}(f)}
	\label{eq:Mydosh}
\end{equation}
The AC susceptibility data provide further evidence for slow magnetic dynamics at low temperature. While the broad feature near 153~K does not show an appreciable frequency shift and is therefore associated with the ferromagnetic transition, the low-temperature anomaly near 30~K in $\chi^{\prime\prime}(T)$ shifts systematically toward higher temperature with increasing frequency, indicating a glassy freezing process. The Mydosh parameter, $\Phi_{f}=0.09$, places the system in the regime of interacting clusters rather than that of a canonical spin glass. To probe the dynamics of the frozen state, the frequency dependence of $T_{f}$ was analyzed using both the Vogel--Fulcher law and the critical slowing-down expression, given in Eqs.~\ref{eq:VF-law} and \ref{eq:CSM}, respectively, following standard analyses of glassy magnetic dynamics and related disordered double perovskites~\cite{souletie1985critical,nair2011griffiths,das2020non}.
\begin{equation}
	\tau = \tau_{0}~exp\left[ \frac{E_{a}}{\kappa_{B}(T-T_{VF})} \right]
	\label{eq:VF-law}
\end{equation}
\begin{equation}
	\tau = \tau_{0}\left( \frac{T_{f}}{T_{g}}-1\right)^{-z\nu}
	\label{eq:CSM}
\end{equation}
The Vogel--Fulcher fit gives $\tau_{0}=8.53\times10^{-4}$~s, $E_{a}/k_{B}=1.96\pm0.07$, and $T_{VF}=27.45\pm0.01$~K, consistent with an interacting cluster ensemble. Likewise, the critical slowing-down analysis yields $\tau_{0}=3.19\times10^{-4}$~s, $z\nu=0.67\pm0.04$, and $T_{g}=28.08\pm0.00$~K. Given the relatively narrow frequency window used in the present experiment, these fitting parameters are best regarded as phenomenological indicators of slow cluster-like magnetic dynamics rather than as definitive critical exponents. Taken together, the AC susceptibility results support slow relaxation and collective freezing of magnetic clusters in GSCM below $\sim 30$~K.

\subsection{Spin-Phonon Coupling}
Raman spectroscopy provides a sensitive probe of local lattice dynamics and symmetry. Room-temperature x-ray diffraction indicates an orthorhombic $Pnma$ average structure for GSCM. The Raman modes are nevertheless discussed using the conventional $P2_{1}/n$ notation commonly adopted for related ordered double perovskites, for which group-theoretical analysis yields 24 Raman-active modes at the $\Gamma$ point, namely 12$A_{g}$ and 12$B_{g}$ modes. In the present disordered sample, these labels are used only as a convenient shorthand for the observed internal stretching and bending vibrations. They are not taken as evidence for a monoclinic average symmetry. Experimentally, only two Raman-active modes are resolved in GSCM.

\begin{figure*}[!t]
	\centering
	\includegraphics[width=0.78\textwidth]{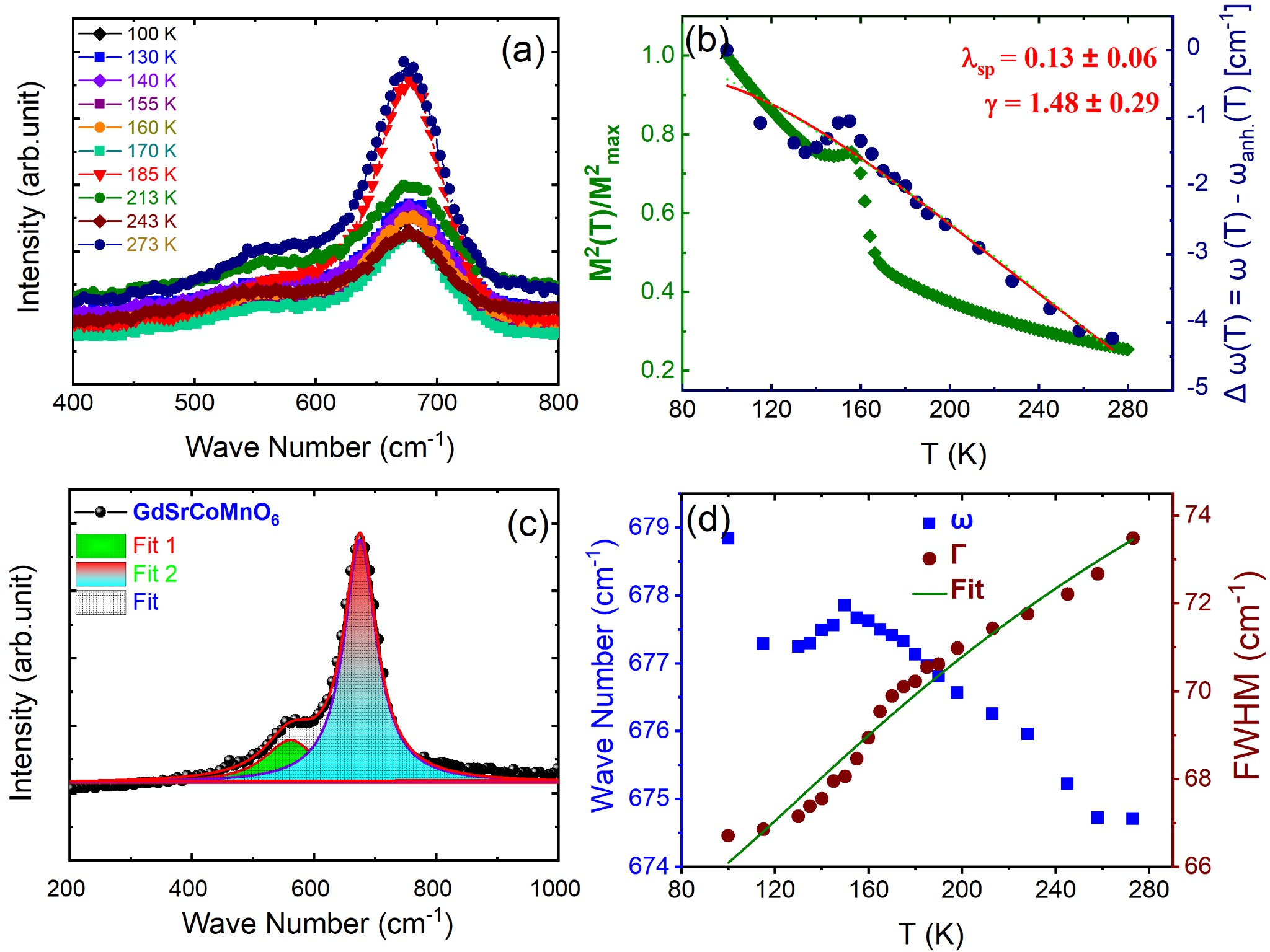}
	\caption{(a) Specific temperature-dependent Raman spectra of GSCM, (b) deconvolution of the Raman spectra and corresponding peak fit at room temperature, (c) anharmonic signature of the Raman peak (672 cm$^{-1}$) as a function of temperature. The absence of spin-phonon coupling in the system is represented by the solid red line. The variation of the full width at half maximum (FWHM) with temperature is displayed on the right-side Y-axis, and (d) shows the reconciliation of Raman spectra with the normalized magnetic moment $M^{2}(T)/M_{max}^{2}$ as a function of temperature.}
	\label{raman}
\end{figure*}

Figure~\ref{raman}(a) shows the temperature evolution of the phonon spectra between 100 and 273~K for a 514~nm excitation line. Two Raman peaks are resolved, in agreement with previous Raman studies of related cobalt-manganite double perovskites~\cite{iliev2007raman,truong2007impact,kumar2014spin,silva2019vibrational,das2019competing}. The higher-frequency peak at about 675.1381~$\pm$~0.17931~$cm^{-1}$ is assigned to the $A_{g}$ stretching mode. The lower-frequency peak at about 561.0828 $\pm$ 1.42386~$cm^{-1}$ is assigned to the $B_{g}$ anti-stretching and bending mode. Low-frequency Raman modes of this type in disordered double perovskites are commonly associated with octahedral tilting and distortion induced by the reduced size of the rare-earth ion~\cite{silva2019vibrational,anshul2020raman}.

The Raman spectra collected over the measured temperature range retain the same overall spectral profile, indicating no obvious change in crystal symmetry across the magnetic transition in GSCM. Lorentzian analysis of the two observed modes reveals the expected temperature-dependent evolution of phonon frequency and linewidth. In particular, the $A_{g}$ stretching mode shows a deviation from the anharmonic behavior on cooling toward the magnetic-ordering region. To examine this feature, the temperature dependence of the phonon frequency was analyzed using the anharmonic model of Balkanski \textit{et al.}~\cite{balkanski1983anharmonic}; the corresponding expressions for the phonon frequency and linewidth are given in Eqs.~\ref{eq:freq} and \ref{eq:linewidth}, respectively. The fit yields $\omega_{0}=677.97\pm0.21$~cm$^{-1}$ and a softening parameter $C=24.02\pm1.26$~cm$^{-1}$. The departure of the experimental phonon position from the anharmonic background near $T_{C}$, together with the accompanying linewidth evolution, suggests an additional contribution associated with spin-lattice coupling.
\begin{equation}
	\omega_{anh}(T)=\omega_{0}-C\left[ 1+\frac{2}{\exp\left( \frac{\hbar \omega_{0}}{2\kappa_{B}T} \right)-1} \right]
	\label{eq:freq}
\end{equation}
\begin{equation}
	\Gamma_{anh}(T)=\Gamma_{0}+\Gamma_{1}\left[ 1+\frac{2}{\exp\left( \frac{\hbar \omega_{0}}{2\kappa_{B}T} \right)-1} \right]
	\label{eq:linewidth}
\end{equation}
where $\Gamma_{1}$ denotes the anharmonic linewidth-broadening coefficient. The anharmonic fit provides the reference needed to evaluate the magnetic contribution to the phonon renormalization.

In magnetic materials, phonon frequencies are sensitive to spin-spin correlations among nearest neighbors. Relating the phonon-mode shift to $\left\langle S_{i}\cdot S_{j} \right\rangle$ reflects the temperature-dependent impact on the spin environment. The strength of spin-phonon coupling can be obtained using Eq.~\ref{eq:strength-SP}
\begin{equation}
 \begin{split}
	\Delta \omega(T) = \omega(T)-\omega_{anh}(T)\approx -\lambda_{sp}S_{i}\cdot S_{j} \\~~
	\approx -4\lambda_{sp}\frac{M^{2}(T)}{\mathrm{M}_{max}^{2}}
	\label{eq:strength-SP}
\end{split}
\end{equation}
where $\omega(T)$ is the renormalized phonon frequency, $\omega_{anh}(T)$ is the anharmonic reference frequency, $\lambda_{sp}$ is the spin-phonon coupling constant, and $M(T)$ is the magnetization measured under the corresponding condition~\cite{granado1999magnetic}. 

Within this framework, the phonon renormalization is related to the spin-correlation function, and the observed $\Delta\omega(T)$ follows the trend of $M^{2}(T)/M_{\max}^{2}$, yielding a spin-phonon coupling constant $\lambda_{sp}\approx0.13$~cm$^{-1}$. These results are therefore consistent with the presence of spin-phonon coupling in GSCM, superimposed on the anharmonic lattice response.

\subsection{Exchange Bias Study}

\begin{figure*}[!t]
    \centering
    \includegraphics[width=0.78\textwidth]{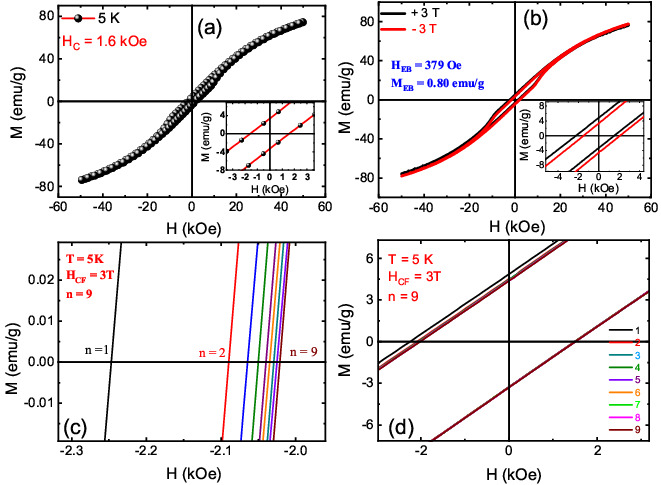}
    \caption{(a) ZFC $M(H)$ loops recorded at 5~K and 50~K, showing no exchange-bias shift. (b) FC $M(H)$ loops recorded at 5~K under cooling fields $H_{CF}=\pm3$~T, showing the exchange-bias effect. The inset shows the enlarged view in the lower-field region. (c) Enlarged view of the left side of the hysteresis curves for different loop indices $n$, and (d) asymmetric training effect recorded for different field cycles at 5~K for GSCM.}
    \label{EB}
\end{figure*}

The shift of the field-cooled $M(H)$ curve relative to the zero-field-cooled $M(H)$ curve towards the negative field axis and positive magnetization axis under a cooling field of ($H_{CF}$) +3~T is shown in Fig.~\ref{EB}(b). A shift in the opposite direction is observed in the field-cooled $M(H)$ curve under a cooling field of ($H_{CF}$) -3~T. This observation confirms the exchange-bias origin of the loop shift.

For comparison, we have also shown the ZFC-$M(H)$ loop recorded at 5~K, and it is found to be symmetric (+$H_{C}$ $\sim$ 1531~Oe and -$H_{C}$ $\sim$ 1531~Oe) about zero field, unlike the FC-$M(H)$ loops that exhibit the exchange-bias effect. However, $M(H)$ does not saturate here due to the presence of competing magnetic phases in the system. The exchange-bias field magnitude, $|H_{EB}|$ [$|H_{EB}|=\left|\frac{H_{C_{1}}+H_{C_{2}}}{2}\right|$ = 379~Oe], and the coercive field [$H_{C}=\frac{\left| H_{C_{1}} \right|+\left|H_{C_{2}} \right|}{2}$ = 1897~Oe] at 5~K are obtained as $H_{C_{1}}$ = -2276~Oe and $H_{C_{2}}$ = 1518~Oe under a +3~T cooling field. In the absence of an exchange-bias shift in the ZFC loops recorded at 5~K and 50~K, field-cooled (FC) magnetization-versus-field ($M$-$H$) measurements were conducted at 5~K after cooling. To examine the asymmetry originating from the FC vertical shift relative to ZFC, the remanent magnetization values of the FC hysteresis loop were employed. This approach used analogous formulas for $M_{EB}$ [$M_{EB}=\frac{M_{C_{1}}+M_{C_{2}}}{2}$] and $M_{C}$ [$M_{C}=\frac{\left| M_{C_{1}} \right|+\left|M_{C_{2}} \right|}{2}$], functioning analogously to $H_{EB}$ and $H_{C}$, respectively~\cite{karmakar2008evidence}. The values of $M_{EB}$ and $M_{C}$ at 3~T were determined to be 0.80~$emu/g$ and 4.1~$emu/g$, respectively, for GSCM. The vertical shift is attributed to pinned interfacial moments, possibly assisted by domain-wall pinning within the FM component.

The significant reduction of the exchange bias ($H_{EB}$) and hysteresis ($H_{C}$) due to continuous field cycling at a particular temperature is known as the training effect. This is a consequence of field-cycle-induced irreversible spin fluctuations at the interface~\cite{paccard1966new}. In the case of the GSCM sample, we observed a decreasing trend in $H_{EB}$ for consecutive hysteresis loops (nine consecutive loops) measured at 5~K under a cooling field of 3~T. A magnified view of the same is shown in Fig.~\ref{EB}(c), indicating the gradual reduction of the exchange-bias parameters under successive field cycling. This behavior can be understood in terms of a rearrangement of the AFM/CG spins and a reduction of the interfacial unidirectional anisotropy. Notice that, in Fig.~\ref{EB}(d), the negative coercivity decays much more rapidly as compared to the positive coercivity within the first field cycle. The training effect is therefore asymmetric and likely reflects the unequal relaxation of the interfacial spin components.

To analyze the training effect, the evolution of $H_{EB}$ with loop index $n$ was examined using three standard descriptions.

The relationship can be expressed by Eq.~\ref{eq:EB} below~\cite{stamps2000mechanisms}
\begin{equation}
	\mathrm{H}_{EB}^{n}-\mathrm{H}_{EB}^{\infty}=\frac{k}{\sqrt{n}}\left( n>1 \right)
	\label{eq:EB}
\end{equation}
The thermal relaxation expression [Eq.~\ref{eq:EB}] provides a reasonable account of the data only for $n>1$, yielding $H_{EB}^{\infty}=109.69\pm1.87$~Oe and $k=220.82\pm0.89$~Oe, but it does not fully reproduce the first-cycle behavior.

\begin{figure*}[!t]
    \centering
    \includegraphics[width=0.78\textwidth]{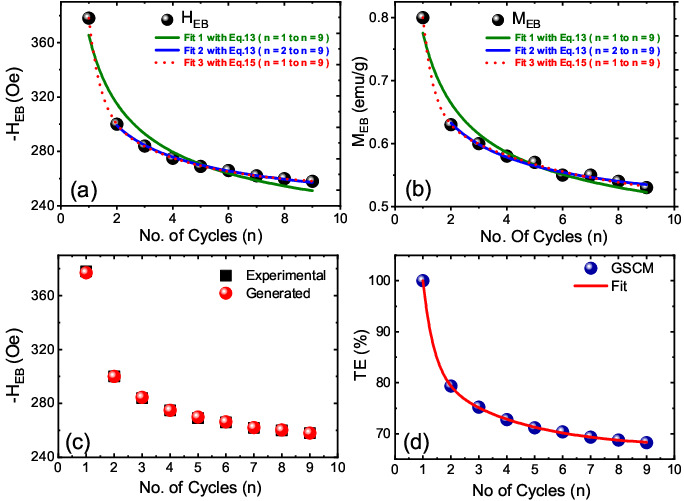}
    \caption{(a) Variation of $H_{EB}$ with loop index $n$ together with the corresponding fits, (b) variation of $M_{EB}$ with loop index $n$, (c) best fit of the Binek model to the experimental data, and (d) training-effect percentage recorded for different field cycles at 5~K for GSCM.}
    \label{EB_fit}
\end{figure*}

Binek~\cite{binek2004training} derived a recursive formula to capture the dependence of $H_{EB}$ on the number of field cycles:
\begin{equation}
	H_{EB}(n+1)-H_{EB}(n)=-\gamma_{H}\left[ H_{EB}(n)-\mathrm{H}_{EB}^{\infty } \right]^{3}
	\label{eq:EB1}
\end{equation}
The Binek recursive model [Eq.~\ref{eq:EB1}] captures the progressive reduction in $H_{EB}$ with cycling more effectively and gives $\gamma_{H}=3.33\pm0.52\times10^{-5}$ and $H_{EB}^{\infty}=220.82\pm0.89$~Oe; the fitted parameters are summarized in Table.~\ref{tab:Eq-14}.

To account for the asymmetry of the training effect more explicitly, the data were further analyzed using the frozen- and rotatable-spin relaxation model [Eq.~\ref{eq:EB2}], which yields $H_{EB}^{\infty}=253.95\pm1.41$~Oe, $A_{f}=1042.59\pm337.73$~Oe, $P_{f}=0.36\pm0.05$, $A_{r}=77.24\pm6.50$~Oe, and $P_{r}=3.11\pm0.35$. The relation $P_{r}>P_{f}$ indicates that the two interfacial spin components relax on different loop scales. Among the models considered, Eq.~\ref{eq:EB2} provides the most comprehensive phenomenological description of the training behavior in GSCM, and the corresponding fitted parameters are summarized in Tables.~\ref{tab:Eq-15a} and \ref{tab:Eq-15b}. Building on this understanding, Mishra et al.~\cite{mishra2009training} introduced the following form:
\begin{equation}
	H_{EB}=H_{E\infty}+A_{f}~exp\left( \frac{-n}{P_{f}} \right)+A_{r}~exp\left( \frac{-n}{P_{r}} \right)
	\label{eq:EB2}
\end{equation}
To quantitatively evaluate the diminishing EB field, we introduced the concept of training-effect percentage (TE$\%$) [Fig.~\ref{EB_fit}(d)] via Eq.~\ref{eq:training}~\cite{ventura2008training}.
\begin{equation}
	TE~\%=\left[ 1-\frac{\left(\mathrm{H}_{EB}^{1}-\mathrm{H}_{EB}^{n} \right)}{\mathrm{H}_{EB}^{1}} \right]~\times 100~\%
	\label{eq:training}
\end{equation}
where $\mathrm{H}_{EB}^{1}$ and $\mathrm{H}_{EB}^{n}$ represent the EB values during the first and $n^{th}$ cycles, respectively. In GSCM, the $\mathrm{H}_{EB}^{2}$ value decreases to 79$\%$ of its initial magnitude [Fig.~\ref{EB_fit}(d)]. Spin rearrangement at the interfaces after each cycle leads to EB-field decay. The TE variation [$\%$] with loop index shows that the EB field in GSCM stabilizes around 68$\%$ of its initial value ($\mathrm{H}_{EB}^{1}$) after the ninth cycle. For clarity, the fitted exchange-bias parameters listed in Tables.~\ref{tab:Eq-13}--\ref{tab:Eq-15b} are presented as absolute magnitudes.

\begin{table*}[!t]
	\centering
	\caption{Parameters extracted by fitting the training-effect data using Eq.~\ref{eq:EB}.}
	\label{tab:Eq-13}
	\begin{tabular}{c c c c c}
		\hline\hline
		$\mathrm{H}_{EB}^{\infty}$~(Oe) & $\kappa_{H}$~(Oe) & $\mathrm{M}_{EB}^{\infty}$~(emu/g) & $\kappa_{M}$~(emu/g) &  \\
		\hline
		109 $\pm$ 1.89 & 220.82 $\pm$ 0.89 & 0.26 $\pm$ 0.01 & 0.44 $\pm$ 0.00 & \\
		\hline\hline
	\end{tabular}
\end{table*}

\begin{table*}[!t]
	\centering
	\caption{Parameters extracted by fitting the training-effect data using Eq.~\ref{eq:EB1}.}
	\label{tab:Eq-14}
	\begin{tabular}{c c c c c}
		\hline\hline
		$\mathrm{H}_{EB}^{\infty}$~(Oe) & $\gamma_{H}$ & $\mathrm{M}_{EB}^{\infty}$~(emu/g) & $\gamma_{M}$ &  \\
		\hline
		220 $\pm$ 0.79 & (3.33 $\pm$ 0.52) $\times$ 10$^{-5}$ & 0.23 $\pm$ 0.06 & (2.1 $\pm$ 0.03) $\times$ 10$^{-4}$ & \\
		\hline\hline
	\end{tabular}
\end{table*}

\begin{table*}[!t]
	\centering
	\caption{Parameters extracted by fitting the training-effect data using Eq.~\ref{eq:EB2}.}
	\label{tab:Eq-15a}
	\begin{tabular}{c c c c c c}
		\hline\hline
		$\mathrm{H}_{EB}^{\infty}$~(Oe) & $A_{f}$~(Oe) & $P_{f}$ & $A_{r}$~(Oe) & $P_{r}$ &  \\
		\hline
		253 $\pm$ 1.41 & 1042.6 $\pm$ 337.73 & 0.36 $\pm$ 0.05 & 77.24 $\pm$ 6.50 & 3.11 $\pm$ 0.35 & \\
		\hline\hline
	\end{tabular}
\end{table*}

\begin{table*}[!t]
	\centering
	\caption{Parameters extracted from fitting the $M_{EB}$ training data using Eq.~\ref{eq:EB2}.}
	\label{tab:Eq-15b}
	\begin{tabular}{c c c c c c}
		\hline\hline
		$\mathrm{M}_{EB}^{\infty}$~(emu/g) & $A_{f}$~(emu/g) & $P_{f}$ & $A_{r}$~(emu/g) & $P_{r}$ &   \\
		\hline
		0.53 $\pm$ 0.02 & 2.28 $\pm$ 1.70 & 0.37 $\pm$ 0.13 & 0.17 $\pm$ 0.01 & 5.04 $\pm$ 2.40 & \\
		\hline\hline
	\end{tabular}
\end{table*}

\subsection{Origin of exchange bias}
\begin{figure*}[!t]
    \centering
    \includegraphics[width=0.7\textwidth]{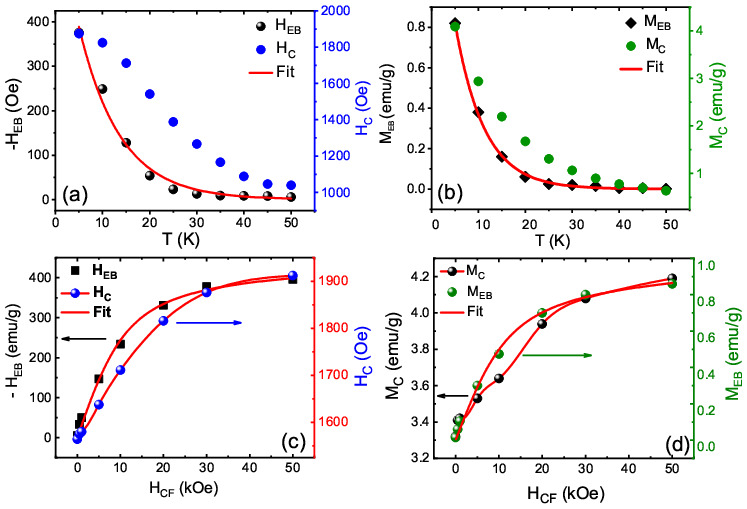}
    \caption{(a) Temperature dependence of $H_{EB}$, (b) temperature dependence of $M_{EB}$, (c) $H_{EB}$ as a function of $H_{CF}$ at 5~K, and (d) $H_{C}$ as a function of $H_{CF}$ at 5~K for GSCM.}
    \label{EB_fit1}
\end{figure*}

The origin of exchange bias (EB) in GSCM is examined through its temperature and cooling-field dependence. On cooling through the low-temperature freezing regime identified by AC susceptibility, the quantities $H_{EB}$, $H_{C}$, $M_{EB}$, and $M_{C}$ all increase [Fig.~\ref{EB_fit1}]. This behavior is consistent with enhanced spin frustration at low temperature~\cite{ventura2008training,huang2008size,moutis2001exchange}. The temperature dependence of the EB parameters is described by the exponential forms given in Eqs.~\ref{eq:EB-fields} and \ref{eq:coercive_field}, shown in Fig.~\ref{EB_fit1}(a) and (b):
\begin{equation}
	H_{EB}=\mathrm{H}_{EB}^{0}~exp\left( \frac{-T}{T_{0}} \right)
	\label{eq:EB-fields}
\end{equation}
\begin{equation}
	M_{EB}=\mathrm{M}_{EB}^{0}~exp\left( \frac{-T}{T^{\ast}} \right)
	\label{eq:coercive_field}
\end{equation}
Here, $T_{0}$ and $T^{\ast}$ are constants, while $\mathrm{H}_{EB}^{0}$ and $\mathrm{M}_{EB}^{0}$ are the extrapolated values of $H_{EB}$ and $M_{EB}$ at absolute zero temperature, respectively. For GSCM, the obtained values are $\mathrm{H}_{EB}^{0}$ = 276.07 $\pm$ 9.40~Oe, $\mathrm{M}_{EB}^{0}$ = 0.47 $\pm$ 0.00~$emu/g$, $T_{0}$ = 11.48 $\pm$ 0.80, and $T^{\ast}$ = 7.48 $\pm$ 0.14.

\begin{figure}[!b]
	\centering
	\includegraphics[width=0.75\columnwidth]{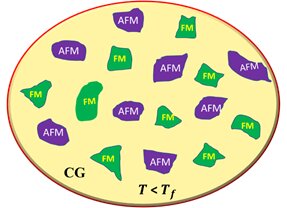}
	\caption{Schematic depiction of the inhomogeneous magnetic phase below the cluster-glass-like (CG) transition in GSCM.}
	\label{schematic}
\end{figure}

The disappearance of EB above 50~K, together with its pronounced enhancement below $T_{f}$, is consistent with the frozen spins in the low-temperature glassy regime contributing strongly to exchange bias in GSCM~\cite{ventura2008training,huang2008size,moutis2001exchange}. The dependence of $H_{EB}$ on cooling field was further analyzed using Eq.~\ref{eq:Heb-Hcool}, which is commonly applied to exchange-biased systems containing ferromagnetic clusters embedded in a magnetically disordered matrix~\cite{niebieskikwiat2005intrinsic}.
\begin{equation}
	H_{EB} \propto~J\left[ \left( \frac{J\mu_{0}}{g\mu_{B}} \right)^{2}~L\left( \frac{\mu~H_{CF}}{\kappa_{B}T_{f}} \right) +H_{CF}\right]
	\label{eq:Heb-Hcool}
\end{equation}
Eq.~\ref{eq:Heb-Hcool} delineates the relationship of $H_{EB}$ with distinct parameters, wherein $J$ signifies the interface exchange constant, $g$ denotes the gyromagnetic factor (with $g$ = 2), $\mu_{B}$ represents the Bohr magneton, $L$ corresponds to the Langevin function, $\mu_{0}$ signifies the magnetic moment per $Co^{3+}$ ion possessing spin $S$ = 1, and $\mu$, which equals N$\mu_{0}$, embodies the magnetic moment of the FM clusters encompassing N spins within the cluster. $T_{f}$~$\approx$~30~K designates the freezing temperature below which both coercivity and EB become markedly enhanced in GdSrCoMnO$_{6}$, although the EB signal remains observable up to 50~K.

Eq.~\ref{eq:Heb-Hcool} has found successful application in assessing the FM cluster size across diverse exchange-biased manganites and cobaltites~\cite{giri2011exchange}. The optimal fit using Eq.~\ref{eq:Heb-Hcool} yields specific parameter values, namely N = 3173 $\pm$ 90 and $J$ = 58 $\pm$ 13~K. Leveraging the acquired N value, the characteristic FM cluster size is estimated to be $D\approx6.31\pm0.33$~nm. This analysis suggests that the exchange-bias response in GSCM is associated with finite ferromagnetic clusters coupled to a magnetically frustrated surrounding phase. Using the low-temperature magnetization scale extracted from the 5~K data, the corresponding cluster density is estimated to be $n\approx6.1\times10^{-4}$~f.u.$^{-1}$. These estimates should be regarded as effective parameters within the adopted model, particularly in view of the additional magnetic background from the Gd$^{3+}$ moments at low temperature.

\subsection{Schematic representation of the EB}
The random distribution of mixed-valent ions (Mn$^{3+}$/Mn$^{4+}$ and Co$^{2+}$/Co$^{3+}$) in GSCM can support the coexistence of FM and AFM regions. The low-temperature cluster-glass-like behavior is then understood as a consequence of the interaction between these competing magnetic components. The exchange-bias effect becomes markedly stronger below $T_{f}$, which suggests that the frozen cluster-glass-like state is important for setting the interfacial anisotropy. The schematic in Fig.~\ref{schematic} illustrates this picture in terms of coexisting FM (green) and AFM (blue) regions. Exchange bias may then arise either from the pinning of cluster-glass-like spins at AFM boundaries or from the anchoring of FM spins to the surrounding frozen cluster-like background. With increasing cooling field $H_{CF}$, the FM regions grow and align more strongly through the Zeeman coupling. Even at 5~T, however, $H_{EB}$ remains unsaturated, which is consistent with persistent AFM regions and strong interfacial frustration.

\section{Summary}
In summary, we have investigated the structural, magnetic, and Raman properties of the disordered double perovskite GdSrCoMnO$_{6}$. The compound crystallizes in a disordered orthorhombic structure and exhibits a ferromagnetic transition at $T_{C} \approx 153$~K. The inverse susceptibility deviates from Curie--Weiss behavior above $T_{C}$ and is consistent with a Griffiths-like regime extending up to $T_{G} \approx 172$~K. Raman measurements show a deviation from the anharmonic phonon background near the magnetic-ordering region, which supports the presence of spin-phonon coupling with $\lambda_{sp}\approx0.13$~cm$^{-1}$.

At low temperature, AC susceptibility indicates slow glassy magnetic dynamics below $T_{f} \approx 30$~K. In the same temperature range, the material exhibits exchange bias that persists up to 50~K, with an exchange-bias magnitude $|H_{EB}| = 379$~Oe at 5~K. The cooling-field dependence of the exchange bias is consistent with an effective ferromagnetic cluster size of $D \approx 6.31 \pm 0.33$~nm. Taken together, these results show that structural disorder, mixed valence, and spin-lattice coupling strongly influence the magnetic ground state and the low-temperature magnetic response of GdSrCoMnO$_{6}$.

\section*{Acknowledgment}
The authors acknowledge the experimental and characterization facilities provided by the Institute of Physics, Bhubaneswar, which were essential for carrying out this work. 

\bibliography{ref}

\begin{thebibliography}{60}%
\makeatletter
\providecommand \@ifxundefined [1]{%
 \@ifx{#1\undefined}
}%
\providecommand \@ifnum [1]{%
 \ifnum #1\expandafter \@firstoftwo
 \else \expandafter \@secondoftwo
 \fi
}%
\providecommand \@ifx [1]{%
 \ifx #1\expandafter \@firstoftwo
 \else \expandafter \@secondoftwo
 \fi
}%
\providecommand \natexlab [1]{#1}%
\providecommand \enquote  [1]{``#1''}%
\providecommand \bibnamefont  [1]{#1}%
\providecommand \bibfnamefont [1]{#1}%
\providecommand \citenamefont [1]{#1}%
\providecommand \href@noop [0]{\@secondoftwo}%
\providecommand \href [0]{\begingroup \@sanitize@url \@href}%
\providecommand \@href[1]{\@@startlink{#1}\@@href}%
\providecommand \@@href[1]{\endgroup#1\@@endlink}%
\providecommand \@sanitize@url [0]{\catcode `\\12\catcode `\$12\catcode
  `\&12\catcode `\#12\catcode `\^12\catcode `\_12\catcode `\%12\relax}%
\providecommand \@@startlink[1]{}%
\providecommand \@@endlink[0]{}%
\providecommand \url  [0]{\begingroup\@sanitize@url \@url }%
\providecommand \@url [1]{\endgroup\@href {#1}{\urlprefix }}%
\providecommand \urlprefix  [0]{URL }%
\providecommand \Eprint [0]{\href }%
\providecommand \doibase [0]{https://doi.org/}%
\providecommand \selectlanguage [0]{\@gobble}%
\providecommand \bibinfo  [0]{\@secondoftwo}%
\providecommand \bibfield  [0]{\@secondoftwo}%
\providecommand \translation [1]{[#1]}%
\providecommand \BibitemOpen [0]{}%
\providecommand \bibitemStop [0]{}%
\providecommand \bibitemNoStop [0]{.\EOS\space}%
\providecommand \EOS [0]{\spacefactor3000\relax}%
\providecommand \BibitemShut  [1]{\csname bibitem#1\endcsname}%
\let\auto@bib@innerbib\@empty
\bibitem [{\citenamefont {Tokura}(2006)}]{tokura2006critical}%
  \BibitemOpen
  \bibfield  {author} {\bibinfo {author} {\bibfnamefont {Y.}~\bibnamefont
  {Tokura}},\ }\bibfield  {title} {\bibinfo {title} {Critical features of
  colossal magnetoresistive manganites},\ }\href
  {https://doi.org/10.1088/0034-4885/69/3/R06} {\bibfield  {journal} {\bibinfo
  {journal} {Rep. Prog. Phys.}\ }\textbf {\bibinfo {volume} {69}},\ \bibinfo
  {pages} {797} (\bibinfo {year} {2006})}\BibitemShut {NoStop}%
\bibitem [{\citenamefont {Mahato}\ \emph {et~al.}(2010)\citenamefont {Mahato},
  \citenamefont {Sethupathi},\ and\ \citenamefont
  {Sankaranarayanan}}]{mahato2010colossal}%
  \BibitemOpen
  \bibfield  {author} {\bibinfo {author} {\bibfnamefont {R.~N.}\ \bibnamefont
  {Mahato}}, \bibinfo {author} {\bibfnamefont {K.}~\bibnamefont {Sethupathi}},\
  and\ \bibinfo {author} {\bibfnamefont {V.}~\bibnamefont {Sankaranarayanan}},\
  }\bibfield  {title} {\bibinfo {title} {Colossal magnetoresistance in the
  double perovskite oxide la$_{2}$comno$_{6}$},\ }\href
  {https://doi.org/10.1063/1.3350907} {\bibfield  {journal} {\bibinfo
  {journal} {J. Appl. Phys.}\ }\textbf {\bibinfo {volume} {107}},\ \bibinfo
  {pages} {09D714} (\bibinfo {year} {2010})}\BibitemShut {NoStop}%
\bibitem [{\citenamefont {Nair}\ \emph {et~al.}(2011)\citenamefont {Nair},
  \citenamefont {Swain}, \citenamefont {Adiga}, \citenamefont {Narayana},\ and\
  \citenamefont {Elizabeth}}]{nair2011griffiths}%
  \BibitemOpen
  \bibfield  {author} {\bibinfo {author} {\bibfnamefont {H.~S.}\ \bibnamefont
  {Nair}}, \bibinfo {author} {\bibfnamefont {D.}~\bibnamefont {Swain}},
  \bibinfo {author} {\bibfnamefont {S.}~\bibnamefont {Adiga}}, \bibinfo
  {author} {\bibfnamefont {C.}~\bibnamefont {Narayana}},\ and\ \bibinfo
  {author} {\bibfnamefont {S.}~\bibnamefont {Elizabeth}},\ }\bibfield  {title}
  {\bibinfo {title} {Griffiths phase-like behavior and spin-phonon coupling in
  double perovskite tb$_{2}$nimno$_{6}$},\ }\href
  {https://doi.org/10.1063/1.3671674} {\bibfield  {journal} {\bibinfo
  {journal} {J. Appl. Phys.}\ }\textbf {\bibinfo {volume} {110}},\ \bibinfo
  {pages} {123919} (\bibinfo {year} {2011})}\BibitemShut {NoStop}%
\bibitem [{\citenamefont {Macedo~Filho}\ \emph {et~al.}(2013)\citenamefont
  {Macedo~Filho}, \citenamefont {Ayala},\ and\ \citenamefont
  {Paschoal}}]{macedo2013spin}%
  \BibitemOpen
  \bibfield  {author} {\bibinfo {author} {\bibfnamefont {R.~B.}\ \bibnamefont
  {Macedo~Filho}}, \bibinfo {author} {\bibfnamefont {A.~P.}\ \bibnamefont
  {Ayala}},\ and\ \bibinfo {author} {\bibfnamefont {C.~W. d.~A.}\ \bibnamefont
  {Paschoal}},\ }\bibfield  {title} {\bibinfo {title} {Spin-phonon coupling in
  y$_{2}$nimno$_{6}$ double perovskite probed by raman spectroscopy},\ }\href
  {https://doi.org/10.1063/1.4804988} {\bibfield  {journal} {\bibinfo
  {journal} {Appl. Phys. Lett.}\ }\textbf {\bibinfo {volume} {102}},\ \bibinfo
  {pages} {192902} (\bibinfo {year} {2013})}\BibitemShut {NoStop}%
\bibitem [{\citenamefont {Zhao}\ \emph {et~al.}(2014)\citenamefont {Zhao},
  \citenamefont {Ren}, \citenamefont {Yang}, \citenamefont {{\'I}{\~n}iguez},
  \citenamefont {Chen},\ and\ \citenamefont {Bellaiche}}]{zhao2014near}%
  \BibitemOpen
  \bibfield  {author} {\bibinfo {author} {\bibfnamefont {H.~J.}\ \bibnamefont
  {Zhao}}, \bibinfo {author} {\bibfnamefont {W.}~\bibnamefont {Ren}}, \bibinfo
  {author} {\bibfnamefont {Y.}~\bibnamefont {Yang}}, \bibinfo {author}
  {\bibfnamefont {J.}~\bibnamefont {{\'I}{\~n}iguez}}, \bibinfo {author}
  {\bibfnamefont {X.~M.}\ \bibnamefont {Chen}},\ and\ \bibinfo {author}
  {\bibfnamefont {L.}~\bibnamefont {Bellaiche}},\ }\bibfield  {title} {\bibinfo
  {title} {Near room-temperature multiferroic materials with tunable
  ferromagnetic and electrical properties},\ }\href
  {https://doi.org/10.1038/ncomms5021} {\bibfield  {journal} {\bibinfo
  {journal} {Nat. Commun.}\ }\textbf {\bibinfo {volume} {5}},\ \bibinfo {pages}
  {4021} (\bibinfo {year} {2014})}\BibitemShut {NoStop}%
\bibitem [{\citenamefont {Su}\ \emph {et~al.}(2015)\citenamefont {Su},
  \citenamefont {Yang}, \citenamefont {Lu}, \citenamefont {Zhang},
  \citenamefont {Gu}, \citenamefont {Lu}, \citenamefont {Li}, \citenamefont
  {Liu},\ and\ \citenamefont {Zhu}}]{su2015magnetism}%
  \BibitemOpen
  \bibfield  {author} {\bibinfo {author} {\bibfnamefont {J.}~\bibnamefont
  {Su}}, \bibinfo {author} {\bibfnamefont {Z.}~\bibnamefont {Yang}}, \bibinfo
  {author} {\bibfnamefont {X.}~\bibnamefont {Lu}}, \bibinfo {author}
  {\bibfnamefont {J.}~\bibnamefont {Zhang}}, \bibinfo {author} {\bibfnamefont
  {L.}~\bibnamefont {Gu}}, \bibinfo {author} {\bibfnamefont {C.}~\bibnamefont
  {Lu}}, \bibinfo {author} {\bibfnamefont {Q.}~\bibnamefont {Li}}, \bibinfo
  {author} {\bibfnamefont {J.-M.}\ \bibnamefont {Liu}},\ and\ \bibinfo {author}
  {\bibfnamefont {J.}~\bibnamefont {Zhu}},\ }\bibfield  {title} {\bibinfo
  {title} {Magnetism-driven ferroelectricity in double perovskite
  y$_{2}$nimno$_{6}$},\ }\href {https://doi.org/10.1021/acsami.5b00911}
  {\bibfield  {journal} {\bibinfo  {journal} {ACS Appl. Mater. Interfaces}\
  }\textbf {\bibinfo {volume} {7}},\ \bibinfo {pages} {13260} (\bibinfo {year}
  {2015})}\BibitemShut {NoStop}%
\bibitem [{\citenamefont {Neenu~Lekshmi}\ and\ \citenamefont
  {Raama~Varma}(2015)}]{neenu2015colossal}%
  \BibitemOpen
  \bibfield  {author} {\bibinfo {author} {\bibfnamefont {P.}~\bibnamefont
  {Neenu~Lekshmi}}\ and\ \bibinfo {author} {\bibfnamefont {M.}~\bibnamefont
  {Raama~Varma}},\ }\bibfield  {title} {\bibinfo {title} {Colossal
  magneto-dielectricity in la$_{2}$nimno$_{6}$ probed by raman spectroscopy},\
  }in\ \href {https://doi.org/10.4028/www.scientific.net/MSF.830-831.513}
  {\emph {\bibinfo {booktitle} {Mater. Sci. Forum}}},\ Vol.\ \bibinfo {volume}
  {830--831}\ (\bibinfo {organization} {Trans Tech Publ},\ \bibinfo {year}
  {2015})\ pp.\ \bibinfo {pages} {513--517}\BibitemShut {NoStop}%
\bibitem [{\citenamefont {Meng}\ \emph {et~al.}(2016)\citenamefont {Meng},
  \citenamefont {Liu}, \citenamefont {Hao}, \citenamefont {Zhang},
  \citenamefont {Yao}, \citenamefont {Meng},\ and\ \citenamefont
  {Zhang}}]{meng2016microscopic}%
  \BibitemOpen
  \bibfield  {author} {\bibinfo {author} {\bibfnamefont {J.}~\bibnamefont
  {Meng}}, \bibinfo {author} {\bibfnamefont {X.}~\bibnamefont {Liu}}, \bibinfo
  {author} {\bibfnamefont {X.}~\bibnamefont {Hao}}, \bibinfo {author}
  {\bibfnamefont {L.}~\bibnamefont {Zhang}}, \bibinfo {author} {\bibfnamefont
  {F.}~\bibnamefont {Yao}}, \bibinfo {author} {\bibfnamefont {J.}~\bibnamefont
  {Meng}},\ and\ \bibinfo {author} {\bibfnamefont {H.}~\bibnamefont {Zhang}},\
  }\bibfield  {title} {\bibinfo {title} {Microscopic mechanistic study on the
  multiferroic of r$_{2}$comno$_{6}$/la$_{2}$comno$_{6}$ (r= ce, pr, nd, pm,
  sm, gd, tb, dy, ho, er, tm) by chemical and hydrostatic pressures: a
  first-principles calculation},\ }\href {https://doi.org/10.1039/C6CP03145E}
  {\bibfield  {journal} {\bibinfo  {journal} {Phys. Chem. Chem. Phys.}\
  }\textbf {\bibinfo {volume} {18}},\ \bibinfo {pages} {23613} (\bibinfo {year}
  {2016})}\BibitemShut {NoStop}%
\bibitem [{\citenamefont {Wang}\ \emph {et~al.}(2015)\citenamefont {Wang},
  \citenamefont {Zhou}, \citenamefont {Wang}, \citenamefont {Cao},
  \citenamefont {Xu},\ and\ \citenamefont {Du}}]{wang2015effect}%
  \BibitemOpen
  \bibfield  {author} {\bibinfo {author} {\bibfnamefont {L.}~\bibnamefont
  {Wang}}, \bibinfo {author} {\bibfnamefont {W.}~\bibnamefont {Zhou}}, \bibinfo
  {author} {\bibfnamefont {D.}~\bibnamefont {Wang}}, \bibinfo {author}
  {\bibfnamefont {Q.}~\bibnamefont {Cao}}, \bibinfo {author} {\bibfnamefont
  {Q.}~\bibnamefont {Xu}},\ and\ \bibinfo {author} {\bibfnamefont
  {Y.}~\bibnamefont {Du}},\ }\bibfield  {title} {\bibinfo {title} {Effect of
  metamagnetism on multiferroic property in double perovskite
  sm$_{2}$comno$_{6}$},\ }\href {https://doi.org/10.1063/1.4917517} {\bibfield
  {journal} {\bibinfo  {journal} {J. Appl. Phys.}\ }\textbf {\bibinfo {volume}
  {117}},\ \bibinfo {pages} {17D914} (\bibinfo {year} {2015})}\BibitemShut
  {NoStop}%
\bibitem [{\citenamefont {Nair}\ \emph {et~al.}(2014)\citenamefont {Nair},
  \citenamefont {Pradheesh}, \citenamefont {Xiao}, \citenamefont {Cherian},
  \citenamefont {Elizabeth}, \citenamefont {Hansen}, \citenamefont
  {Chatterji},\ and\ \citenamefont {Br{\"u}ckel}}]{nair2014magnetization}%
  \BibitemOpen
  \bibfield  {author} {\bibinfo {author} {\bibfnamefont {H.~S.}\ \bibnamefont
  {Nair}}, \bibinfo {author} {\bibfnamefont {R.}~\bibnamefont {Pradheesh}},
  \bibinfo {author} {\bibfnamefont {Y.}~\bibnamefont {Xiao}}, \bibinfo {author}
  {\bibfnamefont {D.}~\bibnamefont {Cherian}}, \bibinfo {author} {\bibfnamefont
  {S.}~\bibnamefont {Elizabeth}}, \bibinfo {author} {\bibfnamefont
  {T.}~\bibnamefont {Hansen}}, \bibinfo {author} {\bibfnamefont
  {T.}~\bibnamefont {Chatterji}},\ and\ \bibinfo {author} {\bibfnamefont
  {T.}~\bibnamefont {Br{\"u}ckel}},\ }\bibfield  {title} {\bibinfo {title}
  {Magnetization-steps in y$_{2}$comno$_{6}$ double perovskite: the role of
  antisite disorder},\ }\href {https://doi.org/10.1063/1.4896399} {\bibfield
  {journal} {\bibinfo  {journal} {J. Appl. Phys.}\ }\textbf {\bibinfo {volume}
  {116}},\ \bibinfo {pages} {123907} (\bibinfo {year} {2014})}\BibitemShut
  {NoStop}%
\bibitem [{\citenamefont {Murthy}\ \emph {et~al.}(2015)\citenamefont {Murthy},
  \citenamefont {Chandrasekhar}, \citenamefont {Mahana}, \citenamefont
  {Topwal},\ and\ \citenamefont {Venimadhav}}]{murthy2015giant}%
  \BibitemOpen
  \bibfield  {author} {\bibinfo {author} {\bibfnamefont {J.~K.}\ \bibnamefont
  {Murthy}}, \bibinfo {author} {\bibfnamefont {K.~D.}\ \bibnamefont
  {Chandrasekhar}}, \bibinfo {author} {\bibfnamefont {S.}~\bibnamefont
  {Mahana}}, \bibinfo {author} {\bibfnamefont {D.}~\bibnamefont {Topwal}},\
  and\ \bibinfo {author} {\bibfnamefont {A.}~\bibnamefont {Venimadhav}},\
  }\bibfield  {title} {\bibinfo {title} {Giant magnetocaloric effect in
  gd$_{2}$nimno$_{6}$ and gd$_{2}$comno$_{6}$ ferromagnetic insulators},\
  }\href {https://doi.org/10.1088/0022-3727/48/35/355001} {\bibfield  {journal}
  {\bibinfo  {journal} {J. Phys. D: Appl. Phys.}\ }\textbf {\bibinfo {volume}
  {48}},\ \bibinfo {pages} {355001} (\bibinfo {year} {2015})}\BibitemShut
  {NoStop}%
\bibitem [{\citenamefont {Singh}\ \emph {et~al.}(2016)\citenamefont {Singh},
  \citenamefont {Chauhan}, \citenamefont {Srivastava},\ and\ \citenamefont
  {Chandra}}]{singh2016influence}%
  \BibitemOpen
  \bibfield  {author} {\bibinfo {author} {\bibfnamefont {A.~K.}\ \bibnamefont
  {Singh}}, \bibinfo {author} {\bibfnamefont {S.}~\bibnamefont {Chauhan}},
  \bibinfo {author} {\bibfnamefont {S.~K.}\ \bibnamefont {Srivastava}},\ and\
  \bibinfo {author} {\bibfnamefont {R.}~\bibnamefont {Chandra}},\ }\bibfield
  {title} {\bibinfo {title} {Influence of antisite disorders on the magnetic
  properties of double perovskite nd$_{2}$nimno$_{6}$},\ }\href
  {https://doi.org/10.1016/j.ssc.2016.04.020} {\bibfield  {journal} {\bibinfo
  {journal} {Solid State Commun.}\ }\textbf {\bibinfo {volume} {242}},\
  \bibinfo {pages} {74} (\bibinfo {year} {2016})}\BibitemShut {NoStop}%
\bibitem [{\citenamefont {Nasir}\ \emph {et~al.}(2019)\citenamefont {Nasir},
  \citenamefont {Kumar}, \citenamefont {Patra}, \citenamefont {Bhattacharya},
  \citenamefont {Jha}, \citenamefont {Basaula}, \citenamefont {Bhatt},
  \citenamefont {Khan}, \citenamefont {Liu}, \citenamefont {Biring} \emph
  {et~al.}}]{nasir2019role}%
  \BibitemOpen
  \bibfield  {author} {\bibinfo {author} {\bibfnamefont {M.}~\bibnamefont
  {Nasir}}, \bibinfo {author} {\bibfnamefont {S.}~\bibnamefont {Kumar}},
  \bibinfo {author} {\bibfnamefont {N.}~\bibnamefont {Patra}}, \bibinfo
  {author} {\bibfnamefont {D.}~\bibnamefont {Bhattacharya}}, \bibinfo {author}
  {\bibfnamefont {S.~N.}\ \bibnamefont {Jha}}, \bibinfo {author} {\bibfnamefont
  {D.~R.}\ \bibnamefont {Basaula}}, \bibinfo {author} {\bibfnamefont
  {S.}~\bibnamefont {Bhatt}}, \bibinfo {author} {\bibfnamefont
  {M.}~\bibnamefont {Khan}}, \bibinfo {author} {\bibfnamefont {S.-W.}\
  \bibnamefont {Liu}}, \bibinfo {author} {\bibfnamefont {S.}~\bibnamefont
  {Biring}}, \emph {et~al.},\ }\bibfield  {title} {\bibinfo {title} {Role of
  antisite disorder, rare-earth size, and superexchange angle on band gap,
  curie temperature, and magnetization of r$_{2}$nimno$_{6}$ double
  perovskites},\ }\href {https://doi.org/10.1021/acsaelm.8b00062} {\bibfield
  {journal} {\bibinfo  {journal} {ACS Appl. Electron. Mater.}\ }\textbf
  {\bibinfo {volume} {1}},\ \bibinfo {pages} {141} (\bibinfo {year}
  {2019})}\BibitemShut {NoStop}%
\bibitem [{\citenamefont {Das}\ \emph {et~al.}(2020)\citenamefont {Das},
  \citenamefont {Sarkar},\ and\ \citenamefont {Mandal}}]{das2020non}%
  \BibitemOpen
  \bibfield  {author} {\bibinfo {author} {\bibfnamefont {M.}~\bibnamefont
  {Das}}, \bibinfo {author} {\bibfnamefont {P.}~\bibnamefont {Sarkar}},\ and\
  \bibinfo {author} {\bibfnamefont {P.}~\bibnamefont {Mandal}},\ }\bibfield
  {title} {\bibinfo {title} {Non-griffiths-like cluster formation in the
  double-perovskite gd$_{2}$comno$_{6}$: Evidence from critical behavior},\
  }\href {https://doi.org/10.1103/PhysRevB.101.144433} {\bibfield  {journal}
  {\bibinfo  {journal} {Phys. Rev. B}\ }\textbf {\bibinfo {volume} {101}},\
  \bibinfo {pages} {144433} (\bibinfo {year} {2020})}\BibitemShut {NoStop}%
\bibitem [{\citenamefont {Salamon}\ \emph {et~al.}(2002)\citenamefont
  {Salamon}, \citenamefont {Lin},\ and\ \citenamefont
  {Chun}}]{salamon2002colossal}%
  \BibitemOpen
  \bibfield  {author} {\bibinfo {author} {\bibfnamefont {M.~B.}\ \bibnamefont
  {Salamon}}, \bibinfo {author} {\bibfnamefont {P.}~\bibnamefont {Lin}},\ and\
  \bibinfo {author} {\bibfnamefont {S.~H.}\ \bibnamefont {Chun}},\ }\bibfield
  {title} {\bibinfo {title} {Colossal magnetoresistance is a griffiths
  singularity},\ }\href {https://doi.org/10.1103/PhysRevLett.88.197203}
  {\bibfield  {journal} {\bibinfo  {journal} {Phys. Rev. Lett.}\ }\textbf
  {\bibinfo {volume} {88}},\ \bibinfo {pages} {197203} (\bibinfo {year}
  {2002})}\BibitemShut {NoStop}%
\bibitem [{\citenamefont {Tong}\ \emph {et~al.}(2008)\citenamefont {Tong},
  \citenamefont {Kim}, \citenamefont {Kwon}, \citenamefont {Qian},
  \citenamefont {Lee}, \citenamefont {Cheong},\ and\ \citenamefont
  {Kim}}]{tong2008griffiths}%
  \BibitemOpen
  \bibfield  {author} {\bibinfo {author} {\bibfnamefont {P.}~\bibnamefont
  {Tong}}, \bibinfo {author} {\bibfnamefont {B.}~\bibnamefont {Kim}}, \bibinfo
  {author} {\bibfnamefont {D.}~\bibnamefont {Kwon}}, \bibinfo {author}
  {\bibfnamefont {T.}~\bibnamefont {Qian}}, \bibinfo {author} {\bibfnamefont
  {S.-I.}\ \bibnamefont {Lee}}, \bibinfo {author} {\bibfnamefont
  {S.}~\bibnamefont {Cheong}},\ and\ \bibinfo {author} {\bibfnamefont {B.~G.}\
  \bibnamefont {Kim}},\ }\bibfield  {title} {\bibinfo {title} {Griffiths phase
  and thermomagnetic irreversibility behavior in slightly electron-doped
  manganites sm$_{1-x}$ca$_{x}$mno$_{3}$ (0.80~$\le$ x $\le$~0.92)},\ }\href
  {https://doi.org/10.1103/PhysRevB.77.184432} {\bibfield  {journal} {\bibinfo
  {journal} {Phys. Rev. B}\ }\textbf {\bibinfo {volume} {77}},\ \bibinfo
  {pages} {184432} (\bibinfo {year} {2008})}\BibitemShut {NoStop}%
\bibitem [{\citenamefont {Jiang}\ \emph {et~al.}(2008)\citenamefont {Jiang},
  \citenamefont {Zhou}, \citenamefont {Williams}, \citenamefont {Mukovskii},\
  and\ \citenamefont {Glazyrin}}]{jiang2008griffiths}%
  \BibitemOpen
  \bibfield  {author} {\bibinfo {author} {\bibfnamefont {W.}~\bibnamefont
  {Jiang}}, \bibinfo {author} {\bibfnamefont {X.}~\bibnamefont {Zhou}},
  \bibinfo {author} {\bibfnamefont {G.}~\bibnamefont {Williams}}, \bibinfo
  {author} {\bibfnamefont {Y.}~\bibnamefont {Mukovskii}},\ and\ \bibinfo
  {author} {\bibfnamefont {K.}~\bibnamefont {Glazyrin}},\ }\bibfield  {title}
  {\bibinfo {title} {Griffiths phase and critical behavior in single-crystal
  la$_{0.7}$ba$_{0.3}$mno$_{3}$: Phase diagram for la$_{1-x}$ba$_{x}$ mno$_{3}$
  ($x \le$ 0.33)},\ }\href {https://doi.org/10.1103/PhysRevB.77.064424}
  {\bibfield  {journal} {\bibinfo  {journal} {Phys. Rev. B}\ }\textbf {\bibinfo
  {volume} {77}},\ \bibinfo {pages} {064424} (\bibinfo {year}
  {2008})}\BibitemShut {NoStop}%
\bibitem [{\citenamefont {Deisenhofer}\ \emph {et~al.}(2005)\citenamefont
  {Deisenhofer}, \citenamefont {Braak}, \citenamefont {Krug~von Nidda},
  \citenamefont {Hemberger}, \citenamefont {Eremina}, \citenamefont {Ivanshin},
  \citenamefont {Balbashov}, \citenamefont {Jug}, \citenamefont {Loidl},
  \citenamefont {Kimura} \emph {et~al.}}]{deisenhofer2005observation}%
  \BibitemOpen
  \bibfield  {author} {\bibinfo {author} {\bibfnamefont {J.}~\bibnamefont
  {Deisenhofer}}, \bibinfo {author} {\bibfnamefont {D.}~\bibnamefont {Braak}},
  \bibinfo {author} {\bibfnamefont {H.-A.}\ \bibnamefont {Krug~von Nidda}},
  \bibinfo {author} {\bibfnamefont {J.}~\bibnamefont {Hemberger}}, \bibinfo
  {author} {\bibfnamefont {R.}~\bibnamefont {Eremina}}, \bibinfo {author}
  {\bibfnamefont {V.}~\bibnamefont {Ivanshin}}, \bibinfo {author}
  {\bibfnamefont {A.}~\bibnamefont {Balbashov}}, \bibinfo {author}
  {\bibfnamefont {G.}~\bibnamefont {Jug}}, \bibinfo {author} {\bibfnamefont
  {A.}~\bibnamefont {Loidl}}, \bibinfo {author} {\bibfnamefont
  {T.}~\bibnamefont {Kimura}}, \emph {et~al.},\ }\bibfield  {title} {\bibinfo
  {title} {Observation of a griffiths phase in paramagnetic
  la$_{1-x}$sr$_{x}$mno$_{3}$},\ }\href
  {https://doi.org/10.1103/PhysRevLett.95.257202} {\bibfield  {journal}
  {\bibinfo  {journal} {Phys. Rev. Lett.}\ }\textbf {\bibinfo {volume} {95}},\
  \bibinfo {pages} {257202} (\bibinfo {year} {2005})}\BibitemShut {NoStop}%
\bibitem [{\citenamefont {Lu}\ \emph {et~al.}(2009)\citenamefont {Lu},
  \citenamefont {Chen}, \citenamefont {Dong}, \citenamefont {Wang},
  \citenamefont {Cai}, \citenamefont {Liu},\ and\ \citenamefont
  {Zhang}}]{lu2009ru}%
  \BibitemOpen
  \bibfield  {author} {\bibinfo {author} {\bibfnamefont {C.~L.}\ \bibnamefont
  {Lu}}, \bibinfo {author} {\bibfnamefont {X.}~\bibnamefont {Chen}}, \bibinfo
  {author} {\bibfnamefont {S.}~\bibnamefont {Dong}}, \bibinfo {author}
  {\bibfnamefont {K.~F.}\ \bibnamefont {Wang}}, \bibinfo {author}
  {\bibfnamefont {H.~L.}\ \bibnamefont {Cai}}, \bibinfo {author} {\bibfnamefont
  {J.-M.}\ \bibnamefont {Liu}},\ and\ \bibinfo {author} {\bibfnamefont {Z.~D.}\
  \bibnamefont {Zhang}},\ }\bibfield  {title} {\bibinfo {title}
  {Ru-doping-induced ferromagnetism in charge-ordered
  la$_{0.4}$ca$_{0.6}$mno$_{3}$},\ }\href
  {https://doi.org/10.1103/PhysRevB.79.245105} {\bibfield  {journal} {\bibinfo
  {journal} {Phys. Rev. B}\ }\textbf {\bibinfo {volume} {79}},\ \bibinfo
  {pages} {245105} (\bibinfo {year} {2009})}\BibitemShut {NoStop}%
\bibitem [{\citenamefont {Qian}\ \emph {et~al.}(2008)\citenamefont {Qian},
  \citenamefont {Tong}, \citenamefont {Kim}, \citenamefont {Lee}, \citenamefont
  {Shin}, \citenamefont {Park},\ and\ \citenamefont
  {Kim}}]{qian2008enhancement}%
  \BibitemOpen
  \bibfield  {author} {\bibinfo {author} {\bibfnamefont {T.}~\bibnamefont
  {Qian}}, \bibinfo {author} {\bibfnamefont {P.}~\bibnamefont {Tong}}, \bibinfo
  {author} {\bibfnamefont {B.}~\bibnamefont {Kim}}, \bibinfo {author}
  {\bibfnamefont {S.-I.}\ \bibnamefont {Lee}}, \bibinfo {author} {\bibfnamefont
  {N.}~\bibnamefont {Shin}}, \bibinfo {author} {\bibfnamefont {S.}~\bibnamefont
  {Park}},\ and\ \bibinfo {author} {\bibfnamefont {B.~G.}\ \bibnamefont
  {Kim}},\ }\bibfield  {title} {\bibinfo {title} {Enhancement of ferromagnetism
  by decreasing tolerance factor in electron-doped manganites},\ }\href
  {https://doi.org/10.1103/PhysRevB.77.094423} {\bibfield  {journal} {\bibinfo
  {journal} {Phys. Rev. B}\ }\textbf {\bibinfo {volume} {77}},\ \bibinfo
  {pages} {094423} (\bibinfo {year} {2008})}\BibitemShut {NoStop}%
\bibitem [{\citenamefont {Yang}\ \emph {et~al.}(2007)\citenamefont {Yang},
  \citenamefont {Sun}, \citenamefont {He}, \citenamefont {Li},\ and\
  \citenamefont {Cheng}}]{yang2007observation}%
  \BibitemOpen
  \bibfield  {author} {\bibinfo {author} {\bibfnamefont {R.-F.}\ \bibnamefont
  {Yang}}, \bibinfo {author} {\bibfnamefont {Y.}~\bibnamefont {Sun}}, \bibinfo
  {author} {\bibfnamefont {W.}~\bibnamefont {He}}, \bibinfo {author}
  {\bibfnamefont {Q.-A.}\ \bibnamefont {Li}},\ and\ \bibinfo {author}
  {\bibfnamefont {Z.-H.}\ \bibnamefont {Cheng}},\ }\bibfield  {title} {\bibinfo
  {title} {Observation of a griffiths-like phase in bilayered manganites},\
  }\href {https://doi.org/10.1063/1.2431061} {\bibfield  {journal} {\bibinfo
  {journal} {Appl. Phys. Lett.}\ }\textbf {\bibinfo {volume} {90}},\ \bibinfo
  {pages} {032502} (\bibinfo {year} {2007})}\BibitemShut {NoStop}%
\bibitem [{\citenamefont {Shimada}\ \emph {et~al.}(2006)\citenamefont
  {Shimada}, \citenamefont {Miyasaka}, \citenamefont {Kumai},\ and\
  \citenamefont {Tokura}}]{shimada2006semiconducting}%
  \BibitemOpen
  \bibfield  {author} {\bibinfo {author} {\bibfnamefont {Y.}~\bibnamefont
  {Shimada}}, \bibinfo {author} {\bibfnamefont {S.}~\bibnamefont {Miyasaka}},
  \bibinfo {author} {\bibfnamefont {R.}~\bibnamefont {Kumai}},\ and\ \bibinfo
  {author} {\bibfnamefont {Y.}~\bibnamefont {Tokura}},\ }\bibfield  {title}
  {\bibinfo {title} {Semiconducting ferromagnetic states in
  la$_{1-x}$sr$_{1+x}$coo$_{4}$},\ }\href
  {https://doi.org/10.1103/PhysRevB.73.134424} {\bibfield  {journal} {\bibinfo
  {journal} {Phys. Rev. B}\ }\textbf {\bibinfo {volume} {73}},\ \bibinfo
  {pages} {134424} (\bibinfo {year} {2006})}\BibitemShut {NoStop}%
\bibitem [{\citenamefont {Liu}\ \emph {et~al.}(2014)\citenamefont {Liu},
  \citenamefont {Shi}, \citenamefont {Zhou}, \citenamefont {Zhao},
  \citenamefont {Li},\ and\ \citenamefont {Guo}}]{liu2014griffiths}%
  \BibitemOpen
  \bibfield  {author} {\bibinfo {author} {\bibfnamefont {W.}~\bibnamefont
  {Liu}}, \bibinfo {author} {\bibfnamefont {L.}~\bibnamefont {Shi}}, \bibinfo
  {author} {\bibfnamefont {S.}~\bibnamefont {Zhou}}, \bibinfo {author}
  {\bibfnamefont {J.}~\bibnamefont {Zhao}}, \bibinfo {author} {\bibfnamefont
  {Y.}~\bibnamefont {Li}},\ and\ \bibinfo {author} {\bibfnamefont
  {Y.}~\bibnamefont {Guo}},\ }\bibfield  {title} {\bibinfo {title} {Griffiths
  phase, spin-phonon coupling, and exchange bias effect in double perovskite
  {Pr$_{2}$CoMnO$_{6}$}},\ }\href {https://doi.org/10.1063/1.4902078}
  {\bibfield  {journal} {\bibinfo  {journal} {J. Appl. Phys.}\ }\textbf
  {\bibinfo {volume} {116}},\ \bibinfo {pages} {193901} (\bibinfo {year}
  {2014})}\BibitemShut {NoStop}%
\bibitem [{\citenamefont {Singh}\ \emph {et~al.}(2018)\citenamefont {Singh},
  \citenamefont {Balasubramanian}, \citenamefont {Singh}, \citenamefont
  {Gupta},\ and\ \citenamefont {Chandra}}]{singh2018structural}%
  \BibitemOpen
  \bibfield  {author} {\bibinfo {author} {\bibfnamefont {A.~K.}\ \bibnamefont
  {Singh}}, \bibinfo {author} {\bibfnamefont {P.}~\bibnamefont
  {Balasubramanian}}, \bibinfo {author} {\bibfnamefont {A.}~\bibnamefont
  {Singh}}, \bibinfo {author} {\bibfnamefont {M.}~\bibnamefont {Gupta}},\ and\
  \bibinfo {author} {\bibfnamefont {R.}~\bibnamefont {Chandra}},\ }\bibfield
  {title} {\bibinfo {title} {Structural transformation, griffiths phase and
  metal-insulator transition in polycrystalline nd$_{2-x}$sr$_{x}$nimno$_{6}$
  (x= 0, 0.2, 0.4, 0.5 and 1) compound},\ }\href
  {https://doi.org/10.1088/1361-648X/aad573} {\bibfield  {journal} {\bibinfo
  {journal} {J. Phys.: Condens. Matter}\ }\textbf {\bibinfo {volume} {30}},\
  \bibinfo {pages} {355401} (\bibinfo {year} {2018})}\BibitemShut {NoStop}%
\bibitem [{\citenamefont {Salamon}\ and\ \citenamefont
  {Chun}(2003)}]{salamon2003griffiths}%
  \BibitemOpen
  \bibfield  {author} {\bibinfo {author} {\bibfnamefont {M.}~\bibnamefont
  {Salamon}}\ and\ \bibinfo {author} {\bibfnamefont {S.}~\bibnamefont {Chun}},\
  }\bibfield  {title} {\bibinfo {title} {Griffiths singularities and
  magnetoresistive manganites},\ }\href
  {https://doi.org/10.1103/PhysRevB.68.014411} {\bibfield  {journal} {\bibinfo
  {journal} {Phys. Rev. B}\ }\textbf {\bibinfo {volume} {68}},\ \bibinfo
  {pages} {014411} (\bibinfo {year} {2003})}\BibitemShut {NoStop}%
\bibitem [{\citenamefont {Castro~Neto}\ \emph {et~al.}(1998)\citenamefont
  {Castro~Neto}, \citenamefont {Castilla},\ and\ \citenamefont
  {Jones}}]{neto1998non}%
  \BibitemOpen
  \bibfield  {author} {\bibinfo {author} {\bibfnamefont {A.~H.}\ \bibnamefont
  {Castro~Neto}}, \bibinfo {author} {\bibfnamefont {G.}~\bibnamefont
  {Castilla}},\ and\ \bibinfo {author} {\bibfnamefont {B.~A.}\ \bibnamefont
  {Jones}},\ }\bibfield  {title} {\bibinfo {title} {Non-fermi liquid behavior
  and griffiths phase in {$f$}-electron compounds},\ }\href
  {https://doi.org/10.1103/PhysRevLett.81.3531} {\bibfield  {journal} {\bibinfo
   {journal} {Phys. Rev. Lett.}\ }\textbf {\bibinfo {volume} {81}},\ \bibinfo
  {pages} {3531} (\bibinfo {year} {1998})}\BibitemShut {NoStop}%
\bibitem [{\citenamefont {Y{\'a}{\~n}ez-Vilar}\ \emph
  {et~al.}(2011)\citenamefont {Y{\'a}{\~n}ez-Vilar}, \citenamefont {Mun},
  \citenamefont {Zapf}, \citenamefont {Ueland}, \citenamefont {Gardner},
  \citenamefont {Thompson}, \citenamefont {Singleton}, \citenamefont
  {S{\'a}nchez-And{\'u}jar}, \citenamefont {Mira}, \citenamefont {Biskup} \emph
  {et~al.}}]{yanez2011multiferroic}%
  \BibitemOpen
  \bibfield  {author} {\bibinfo {author} {\bibfnamefont {S.}~\bibnamefont
  {Y{\'a}{\~n}ez-Vilar}}, \bibinfo {author} {\bibfnamefont {E.~D.}\
  \bibnamefont {Mun}}, \bibinfo {author} {\bibfnamefont {V.~S.}\ \bibnamefont
  {Zapf}}, \bibinfo {author} {\bibfnamefont {B.~G.}\ \bibnamefont {Ueland}},
  \bibinfo {author} {\bibfnamefont {J.~S.}\ \bibnamefont {Gardner}}, \bibinfo
  {author} {\bibfnamefont {J.~D.}\ \bibnamefont {Thompson}}, \bibinfo {author}
  {\bibfnamefont {J.}~\bibnamefont {Singleton}}, \bibinfo {author}
  {\bibfnamefont {M.}~\bibnamefont {S{\'a}nchez-And{\'u}jar}}, \bibinfo
  {author} {\bibfnamefont {J.}~\bibnamefont {Mira}}, \bibinfo {author}
  {\bibfnamefont {N.}~\bibnamefont {Biskup}}, \emph {et~al.},\ }\bibfield
  {title} {\bibinfo {title} {Multiferroic behavior in the double-perovskite
  lu$_{2}$mncoo$_{6}$},\ }\href {https://doi.org/10.1103/PhysRevB.84.134427}
  {\bibfield  {journal} {\bibinfo  {journal} {Phys. Rev. B}\ }\textbf {\bibinfo
  {volume} {84}},\ \bibinfo {pages} {134427} (\bibinfo {year}
  {2011})}\BibitemShut {NoStop}%
\bibitem [{\citenamefont {Chikara}\ \emph {et~al.}(2016)\citenamefont
  {Chikara}, \citenamefont {Singleton}, \citenamefont {Bowlan}, \citenamefont
  {Yarotski}, \citenamefont {Lee}, \citenamefont {Choi}, \citenamefont {Choi},\
  and\ \citenamefont {Zapf}}]{chikara2016electric}%
  \BibitemOpen
  \bibfield  {author} {\bibinfo {author} {\bibfnamefont {S.}~\bibnamefont
  {Chikara}}, \bibinfo {author} {\bibfnamefont {J.}~\bibnamefont {Singleton}},
  \bibinfo {author} {\bibfnamefont {J.}~\bibnamefont {Bowlan}}, \bibinfo
  {author} {\bibfnamefont {D.~A.}\ \bibnamefont {Yarotski}}, \bibinfo {author}
  {\bibfnamefont {N.}~\bibnamefont {Lee}}, \bibinfo {author} {\bibfnamefont
  {H.~Y.}\ \bibnamefont {Choi}}, \bibinfo {author} {\bibfnamefont {Y.~J.}\
  \bibnamefont {Choi}},\ and\ \bibinfo {author} {\bibfnamefont {V.~S.}\
  \bibnamefont {Zapf}},\ }\bibfield  {title} {\bibinfo {title} {Electric
  polarization observed in single crystals of multiferroic
  lu$_{2}$mncoo$_{6}$},\ }\href {https://doi.org/10.1103/PhysRevB.93.180405}
  {\bibfield  {journal} {\bibinfo  {journal} {Phys. Rev. B}\ }\textbf {\bibinfo
  {volume} {93}},\ \bibinfo {pages} {180405} (\bibinfo {year}
  {2016})}\BibitemShut {NoStop}%
\bibitem [{\citenamefont {Bhatti}\ \emph {et~al.}(2019)\citenamefont {Bhatti},
  \citenamefont {Bhatti}, \citenamefont {Mahato},\ and\ \citenamefont
  {Ahsan}}]{bhatti2019magnetic}%
  \BibitemOpen
  \bibfield  {author} {\bibinfo {author} {\bibfnamefont {I.~N.}\ \bibnamefont
  {Bhatti}}, \bibinfo {author} {\bibfnamefont {I.~N.}\ \bibnamefont {Bhatti}},
  \bibinfo {author} {\bibfnamefont {R.~N.}\ \bibnamefont {Mahato}},\ and\
  \bibinfo {author} {\bibfnamefont {M.~A.~H.}\ \bibnamefont {Ahsan}},\
  }\bibfield  {title} {\bibinfo {title} {Magnetic behavior, griffiths phase and
  magneto-transport study in 3$d$ based nano-crystalline double perovskite
  pr$_{2}$comno$_{6}$},\ }\href
  {https://doi.org/10.1016/j.physleta.2019.04.036} {\bibfield  {journal}
  {\bibinfo  {journal} {Phys. Lett. A}\ }\textbf {\bibinfo {volume} {383}},\
  \bibinfo {pages} {2326} (\bibinfo {year} {2019})}\BibitemShut {NoStop}%
\bibitem [{\citenamefont {Choudhury}\ \emph {et~al.}(2012)\citenamefont
  {Choudhury}, \citenamefont {Mandal}, \citenamefont {Mathieu}, \citenamefont
  {Hazarika}, \citenamefont {Rajan}, \citenamefont {Sundaresan}, \citenamefont
  {Waghmare}, \citenamefont {Knut}, \citenamefont {Karis}, \citenamefont
  {Nordblad} \emph {et~al.}}]{choudhury2012near}%
  \BibitemOpen
  \bibfield  {author} {\bibinfo {author} {\bibfnamefont {D.}~\bibnamefont
  {Choudhury}}, \bibinfo {author} {\bibfnamefont {P.}~\bibnamefont {Mandal}},
  \bibinfo {author} {\bibfnamefont {R.}~\bibnamefont {Mathieu}}, \bibinfo
  {author} {\bibfnamefont {A.}~\bibnamefont {Hazarika}}, \bibinfo {author}
  {\bibfnamefont {S.}~\bibnamefont {Rajan}}, \bibinfo {author} {\bibfnamefont
  {A.}~\bibnamefont {Sundaresan}}, \bibinfo {author} {\bibfnamefont
  {U.}~\bibnamefont {Waghmare}}, \bibinfo {author} {\bibfnamefont
  {R.}~\bibnamefont {Knut}}, \bibinfo {author} {\bibfnamefont {O.}~\bibnamefont
  {Karis}}, \bibinfo {author} {\bibfnamefont {P.}~\bibnamefont {Nordblad}},
  \emph {et~al.},\ }\bibfield  {title} {\bibinfo {title} {Near-room-temperature
  colossal magnetodielectricity and multiglass properties in partially
  disordered la$_{2}$nimno$_{6}$},\ }\href
  {https://doi.org/10.1103/PhysRevLett.108.127201} {\bibfield  {journal}
  {\bibinfo  {journal} {Phys. Rev. Lett.}\ }\textbf {\bibinfo {volume} {108}},\
  \bibinfo {pages} {127201} (\bibinfo {year} {2012})}\BibitemShut {NoStop}%
\bibitem [{\citenamefont {Banerjee}\ \emph {et~al.}(2018)\citenamefont
  {Banerjee}, \citenamefont {Sannigrahi}, \citenamefont {Giri},\ and\
  \citenamefont {Majumdar}}]{banerjee2018magnetization}%
  \BibitemOpen
  \bibfield  {author} {\bibinfo {author} {\bibfnamefont {A.}~\bibnamefont
  {Banerjee}}, \bibinfo {author} {\bibfnamefont {J.}~\bibnamefont
  {Sannigrahi}}, \bibinfo {author} {\bibfnamefont {S.}~\bibnamefont {Giri}},\
  and\ \bibinfo {author} {\bibfnamefont {S.}~\bibnamefont {Majumdar}},\
  }\bibfield  {title} {\bibinfo {title} {Magnetization reversal and inverse
  exchange bias phenomenon in the ferrimagnetic polycrystalline compound
  er$_{2}$comno$_{6}$},\ }\href {https://doi.org/10.1103/PhysRevB.98.104414}
  {\bibfield  {journal} {\bibinfo  {journal} {Phys. Rev. B}\ }\textbf {\bibinfo
  {volume} {98}},\ \bibinfo {pages} {104414} (\bibinfo {year}
  {2018})}\BibitemShut {NoStop}%
\bibitem [{\citenamefont {Madhogaria}\ \emph {et~al.}(2019)\citenamefont
  {Madhogaria}, \citenamefont {Das}, \citenamefont {Clements}, \citenamefont
  {Kalappattil}, \citenamefont {Phan}, \citenamefont {Srikanth}, \citenamefont
  {Dang}, \citenamefont {Kozlenko},\ and\ \citenamefont
  {Bingham}}]{madhogaria2019evidence}%
  \BibitemOpen
  \bibfield  {author} {\bibinfo {author} {\bibfnamefont {R.}~\bibnamefont
  {Madhogaria}}, \bibinfo {author} {\bibfnamefont {R.}~\bibnamefont {Das}},
  \bibinfo {author} {\bibfnamefont {E.}~\bibnamefont {Clements}}, \bibinfo
  {author} {\bibfnamefont {V.}~\bibnamefont {Kalappattil}}, \bibinfo {author}
  {\bibfnamefont {M.}~\bibnamefont {Phan}}, \bibinfo {author} {\bibfnamefont
  {H.}~\bibnamefont {Srikanth}}, \bibinfo {author} {\bibfnamefont
  {N.}~\bibnamefont {Dang}}, \bibinfo {author} {\bibfnamefont {D.}~\bibnamefont
  {Kozlenko}},\ and\ \bibinfo {author} {\bibfnamefont {N.}~\bibnamefont
  {Bingham}},\ }\bibfield  {title} {\bibinfo {title} {Evidence of long-range
  ferromagnetic order and spin frustration effects in the double perovskite
  la$_{2}$comno$_{6}$},\ }\href {https://doi.org/10.1103/PhysRevB.99.104436}
  {\bibfield  {journal} {\bibinfo  {journal} {Phys. Rev. B}\ }\textbf {\bibinfo
  {volume} {99}},\ \bibinfo {pages} {104436} (\bibinfo {year}
  {2019})}\BibitemShut {NoStop}%
\bibitem [{\citenamefont {Blasco}\ \emph {et~al.}(2015)\citenamefont {Blasco},
  \citenamefont {Garc{\'i}a-Mu{\~n}oz}, \citenamefont {Garc{\'i}a},
  \citenamefont {Stankiewicz}, \citenamefont {Sub{\'i}as}, \citenamefont
  {Ritter},\ and\ \citenamefont
  {Rodr{\'i}guez-Velamaz{\'a}n}}]{blasco2015evidence}%
  \BibitemOpen
  \bibfield  {author} {\bibinfo {author} {\bibfnamefont {J.}~\bibnamefont
  {Blasco}}, \bibinfo {author} {\bibfnamefont {J.~L.}\ \bibnamefont
  {Garc{\'i}a-Mu{\~n}oz}}, \bibinfo {author} {\bibfnamefont {J.}~\bibnamefont
  {Garc{\'i}a}}, \bibinfo {author} {\bibfnamefont {J.}~\bibnamefont
  {Stankiewicz}}, \bibinfo {author} {\bibfnamefont {G.}~\bibnamefont
  {Sub{\'i}as}}, \bibinfo {author} {\bibfnamefont {C.}~\bibnamefont {Ritter}},\
  and\ \bibinfo {author} {\bibfnamefont {J.~A.}\ \bibnamefont
  {Rodr{\'i}guez-Velamaz{\'a}n}},\ }\bibfield  {title} {\bibinfo {title}
  {Evidence of large magneto-dielectric effect coupled to a metamagnetic
  transition in yb$_{2}$comno$_{6}$},\ }\href
  {https://doi.org/10.1063/1.4926403} {\bibfield  {journal} {\bibinfo
  {journal} {Appl. Phys. Lett.}\ }\textbf {\bibinfo {volume} {107}},\ \bibinfo
  {pages} {012902} (\bibinfo {year} {2015})}\BibitemShut {NoStop}%
\bibitem [{\citenamefont {Moon}\ \emph {et~al.}(2018)\citenamefont {Moon},
  \citenamefont {Kim}, \citenamefont {Oh}, \citenamefont {Kim}, \citenamefont
  {Shin}, \citenamefont {Choi},\ and\ \citenamefont
  {Lee}}]{moon2018anisotropic}%
  \BibitemOpen
  \bibfield  {author} {\bibinfo {author} {\bibfnamefont {J.}~\bibnamefont
  {Moon}}, \bibinfo {author} {\bibfnamefont {M.}~\bibnamefont {Kim}}, \bibinfo
  {author} {\bibfnamefont {D.}~\bibnamefont {Oh}}, \bibinfo {author}
  {\bibfnamefont {J.}~\bibnamefont {Kim}}, \bibinfo {author} {\bibfnamefont
  {H.}~\bibnamefont {Shin}}, \bibinfo {author} {\bibfnamefont {Y.}~\bibnamefont
  {Choi}},\ and\ \bibinfo {author} {\bibfnamefont {N.}~\bibnamefont {Lee}},\
  }\bibfield  {title} {\bibinfo {title} {Anisotropic magnetic properties and
  giant rotating magnetocaloric effect in double-perovskite
  tb$_{2}$comno$_{6}$},\ }\href {https://doi.org/10.1103/PhysRevB.98.174424}
  {\bibfield  {journal} {\bibinfo  {journal} {Phys. Rev. B}\ }\textbf {\bibinfo
  {volume} {98}},\ \bibinfo {pages} {174424} (\bibinfo {year}
  {2018})}\BibitemShut {NoStop}%
\bibitem [{\citenamefont {Sahoo}\ \emph {et~al.}(2019)\citenamefont {Sahoo},
  \citenamefont {Takeuchi}, \citenamefont {Ohtomo},\ and\ \citenamefont
  {Hossain}}]{sahoo2019exchange}%
  \BibitemOpen
  \bibfield  {author} {\bibinfo {author} {\bibfnamefont {R.}~\bibnamefont
  {Sahoo}}, \bibinfo {author} {\bibfnamefont {Y.}~\bibnamefont {Takeuchi}},
  \bibinfo {author} {\bibfnamefont {A.}~\bibnamefont {Ohtomo}},\ and\ \bibinfo
  {author} {\bibfnamefont {Z.}~\bibnamefont {Hossain}},\ }\bibfield  {title}
  {\bibinfo {title} {Exchange bias and spin glass states driven by antisite
  disorder in the double perovskite compound lasrcofeo$_{6}$},\ }\href
  {https://doi.org/10.1103/PhysRevB.100.214436} {\bibfield  {journal} {\bibinfo
   {journal} {Phys. Rev. B}\ }\textbf {\bibinfo {volume} {100}},\ \bibinfo
  {pages} {214436} (\bibinfo {year} {2019})}\BibitemShut {NoStop}%
\bibitem [{\citenamefont {Silva~Jr}\ \emph {et~al.}(2022)\citenamefont
  {Silva~Jr}, \citenamefont {Santos}, \citenamefont {Escote}, \citenamefont
  {Costa}, \citenamefont {Moreno}, \citenamefont {Paz}, \citenamefont
  {Ang{\'e}lica},\ and\ \citenamefont {Ferreira}}]{silva2022griffiths}%
  \BibitemOpen
  \bibfield  {author} {\bibinfo {author} {\bibfnamefont {R.}~\bibnamefont
  {Silva~Jr}}, \bibinfo {author} {\bibfnamefont {C.}~\bibnamefont {Santos}},
  \bibinfo {author} {\bibfnamefont {M.}~\bibnamefont {Escote}}, \bibinfo
  {author} {\bibfnamefont {B.}~\bibnamefont {Costa}}, \bibinfo {author}
  {\bibfnamefont {N.}~\bibnamefont {Moreno}}, \bibinfo {author} {\bibfnamefont
  {S.}~\bibnamefont {Paz}}, \bibinfo {author} {\bibfnamefont {R.}~\bibnamefont
  {Ang{\'e}lica}},\ and\ \bibinfo {author} {\bibfnamefont {N.}~\bibnamefont
  {Ferreira}},\ }\bibfield  {title} {\bibinfo {title} {Griffiths-like phase,
  large magnetocaloric effect, and unconventional critical behavior in the
  ndsrcofeo$_{6}$ disordered double perovskite},\ }\href
  {https://doi.org/10.1103/PhysRevB.106.134439} {\bibfield  {journal} {\bibinfo
   {journal} {Phys. Rev. B}\ }\textbf {\bibinfo {volume} {106}},\ \bibinfo
  {pages} {134439} (\bibinfo {year} {2022})}\BibitemShut {NoStop}%
\bibitem [{\citenamefont {Silva~Jr}\ \emph {et~al.}(2023)\citenamefont
  {Silva~Jr}, \citenamefont {Gainza}, \citenamefont {dos Santos}, \citenamefont
  {Rodrigues}, \citenamefont {Dejoie}, \citenamefont {Huttel}, \citenamefont
  {Biskup}, \citenamefont {Nemes}, \citenamefont {Mart{\'\i}nez}, \citenamefont
  {Ferreira} \emph {et~al.}}]{silva2023magnetoelastic}%
  \BibitemOpen
  \bibfield  {author} {\bibinfo {author} {\bibfnamefont {R.~S.}\ \bibnamefont
  {Silva~Jr}}, \bibinfo {author} {\bibfnamefont {J.}~\bibnamefont {Gainza}},
  \bibinfo {author} {\bibfnamefont {C.}~\bibnamefont {dos Santos}}, \bibinfo
  {author} {\bibfnamefont {J.~E.~F.}\ \bibnamefont {Rodrigues}}, \bibinfo
  {author} {\bibfnamefont {C.}~\bibnamefont {Dejoie}}, \bibinfo {author}
  {\bibfnamefont {Y.}~\bibnamefont {Huttel}}, \bibinfo {author} {\bibfnamefont
  {N.}~\bibnamefont {Biskup}}, \bibinfo {author} {\bibfnamefont {N.~M.}\
  \bibnamefont {Nemes}}, \bibinfo {author} {\bibfnamefont {J.~L.}\ \bibnamefont
  {Mart{\'\i}nez}}, \bibinfo {author} {\bibfnamefont {N.~S.}\ \bibnamefont
  {Ferreira}}, \emph {et~al.},\ }\bibfield  {title} {\bibinfo {title}
  {Magnetoelastic coupling and cryogenic magnetocaloric effect in two-site
  disordered {GdSrCoFeO$_{6}$} double perovskite},\ }\href
  {https://doi.org/10.1021/acs.chemmater.2c03574} {\bibfield  {journal}
  {\bibinfo  {journal} {Chem. Mater.}\ }\textbf {\bibinfo {volume} {35}},\
  \bibinfo {pages} {2439} (\bibinfo {year} {2023})}\BibitemShut {NoStop}%
\bibitem [{\citenamefont {Moon}\ \emph {et~al.}(2017)\citenamefont {Moon},
  \citenamefont {Kim}, \citenamefont {Choi},\ and\ \citenamefont
  {Lee}}]{moon2017giant}%
  \BibitemOpen
  \bibfield  {author} {\bibinfo {author} {\bibfnamefont {J.~Y.}\ \bibnamefont
  {Moon}}, \bibinfo {author} {\bibfnamefont {M.~K.}\ \bibnamefont {Kim}},
  \bibinfo {author} {\bibfnamefont {Y.~J.}\ \bibnamefont {Choi}},\ and\
  \bibinfo {author} {\bibfnamefont {N.}~\bibnamefont {Lee}},\ }\bibfield
  {title} {\bibinfo {title} {Giant anisotropic magnetocaloric effect in
  double-perovskite gd$_{2}$comno$_{6}$ single crystals},\ }\href
  {https://doi.org/10.1038/s41598-017-16416-z} {\bibfield  {journal} {\bibinfo
  {journal} {Sci. Rep.}\ }\textbf {\bibinfo {volume} {7}},\ \bibinfo {pages}
  {16099} (\bibinfo {year} {2017})}\BibitemShut {NoStop}%
\bibitem [{\citenamefont {Wu}\ and\ \citenamefont
  {Leighton}(2003)}]{wu2003glassy}%
  \BibitemOpen
  \bibfield  {author} {\bibinfo {author} {\bibfnamefont {J.}~\bibnamefont
  {Wu}}\ and\ \bibinfo {author} {\bibfnamefont {C.}~\bibnamefont {Leighton}},\
  }\bibfield  {title} {\bibinfo {title} {Glassy ferromagnetism and magnetic
  phase separation in la$_{1-x}$sr$_{x}$coo$_{3}$},\ }\href
  {https://doi.org/10.1103/PhysRevB.67.174408} {\bibfield  {journal} {\bibinfo
  {journal} {Phys. Rev. B}\ }\textbf {\bibinfo {volume} {67}},\ \bibinfo
  {pages} {174408} (\bibinfo {year} {2003})}\BibitemShut {NoStop}%
\bibitem [{\citenamefont {Bull}\ \emph {et~al.}(2016)\citenamefont {Bull},
  \citenamefont {Playford}, \citenamefont {Knight}, \citenamefont {Stenning},\
  and\ \citenamefont {Tucker}}]{bull2016magnetic}%
  \BibitemOpen
  \bibfield  {author} {\bibinfo {author} {\bibfnamefont {C.}~\bibnamefont
  {Bull}}, \bibinfo {author} {\bibfnamefont {H.}~\bibnamefont {Playford}},
  \bibinfo {author} {\bibfnamefont {K.}~\bibnamefont {Knight}}, \bibinfo
  {author} {\bibfnamefont {G.}~\bibnamefont {Stenning}},\ and\ \bibinfo
  {author} {\bibfnamefont {M.}~\bibnamefont {Tucker}},\ }\bibfield  {title}
  {\bibinfo {title} {Magnetic and structural phase diagram of the solid
  solution laco$_{x}$mn$_{1-x}$o$_{3}$},\ }\href
  {https://doi.org/10.1103/PhysRevB.94.014102} {\bibfield  {journal} {\bibinfo
  {journal} {Phys. Rev. B}\ }\textbf {\bibinfo {volume} {94}},\ \bibinfo
  {pages} {014102} (\bibinfo {year} {2016})}\BibitemShut {NoStop}%
\bibitem [{\citenamefont {Mydosh}(1993)}]{mydosh1993spin}%
  \BibitemOpen
  \bibfield  {author} {\bibinfo {author} {\bibfnamefont {J.~A.}\ \bibnamefont
  {Mydosh}},\ }\href@noop {} {\emph {\bibinfo {title} {Spin Glasses: An
  Experimental Introduction}}}\ (\bibinfo  {publisher} {Taylor \& Francis},\
  \bibinfo {address} {London},\ \bibinfo {year} {1993})\BibitemShut {NoStop}%
\bibitem [{\citenamefont {Souletie}\ and\ \citenamefont
  {Tholence}(1985)}]{souletie1985critical}%
  \BibitemOpen
  \bibfield  {author} {\bibinfo {author} {\bibfnamefont {J.}~\bibnamefont
  {Souletie}}\ and\ \bibinfo {author} {\bibfnamefont {J.}~\bibnamefont
  {Tholence}},\ }\bibfield  {title} {\bibinfo {title} {Critical slowing down in
  spin glasses and other glasses: Fulcher versus power law},\ }\href
  {https://doi.org/10.1103/PhysRevB.32.516} {\bibfield  {journal} {\bibinfo
  {journal} {Phys. Rev. B}\ }\textbf {\bibinfo {volume} {32}},\ \bibinfo
  {pages} {516} (\bibinfo {year} {1985})}\BibitemShut {NoStop}%
\bibitem [{\citenamefont {Iliev}\ \emph {et~al.}(2007)\citenamefont {Iliev},
  \citenamefont {Abrashev}, \citenamefont {Litvinchuk}, \citenamefont
  {Hadjiev}, \citenamefont {Guo},\ and\ \citenamefont
  {Gupta}}]{iliev2007raman}%
  \BibitemOpen
  \bibfield  {author} {\bibinfo {author} {\bibfnamefont {M.}~\bibnamefont
  {Iliev}}, \bibinfo {author} {\bibfnamefont {M.}~\bibnamefont {Abrashev}},
  \bibinfo {author} {\bibfnamefont {A.}~\bibnamefont {Litvinchuk}}, \bibinfo
  {author} {\bibfnamefont {V.}~\bibnamefont {Hadjiev}}, \bibinfo {author}
  {\bibfnamefont {H.}~\bibnamefont {Guo}},\ and\ \bibinfo {author}
  {\bibfnamefont {A.}~\bibnamefont {Gupta}},\ }\bibfield  {title} {\bibinfo
  {title} {Raman spectroscopy of ordered double perovskite la$_{2}$comno$_{6}$
  thin films},\ }\href {https://doi.org/10.1103/PhysRevB.75.104118} {\bibfield
  {journal} {\bibinfo  {journal} {Phys. Rev. B}\ }\textbf {\bibinfo {volume}
  {75}},\ \bibinfo {pages} {104118} (\bibinfo {year} {2007})}\BibitemShut
  {NoStop}%
\bibitem [{\citenamefont {Truong}\ \emph {et~al.}(2007)\citenamefont {Truong},
  \citenamefont {Laverdi{\`e}re}, \citenamefont {Singh}, \citenamefont
  {Jandl},\ and\ \citenamefont {Fournier}}]{truong2007impact}%
  \BibitemOpen
  \bibfield  {author} {\bibinfo {author} {\bibfnamefont {K.}~\bibnamefont
  {Truong}}, \bibinfo {author} {\bibfnamefont {J.}~\bibnamefont
  {Laverdi{\`e}re}}, \bibinfo {author} {\bibfnamefont {M.}~\bibnamefont
  {Singh}}, \bibinfo {author} {\bibfnamefont {S.}~\bibnamefont {Jandl}},\ and\
  \bibinfo {author} {\bibfnamefont {P.}~\bibnamefont {Fournier}},\ }\bibfield
  {title} {\bibinfo {title} {Impact of co$/$ mn cation ordering on phonon
  anomalies in la$_{2}$comno$_{6}$ double perovskites: Raman spectroscopy},\
  }\href {https://doi.org/10.1103/PhysRevB.76.132413} {\bibfield  {journal}
  {\bibinfo  {journal} {Phys. Rev. B}\ }\textbf {\bibinfo {volume} {76}},\
  \bibinfo {pages} {132413} (\bibinfo {year} {2007})}\BibitemShut {NoStop}%
\bibitem [{\citenamefont {Kumar}\ \emph {et~al.}(2014)\citenamefont {Kumar},
  \citenamefont {Kumar},\ and\ \citenamefont {Sathe}}]{kumar2014spin}%
  \BibitemOpen
  \bibfield  {author} {\bibinfo {author} {\bibfnamefont {D.}~\bibnamefont
  {Kumar}}, \bibinfo {author} {\bibfnamefont {S.}~\bibnamefont {Kumar}},\ and\
  \bibinfo {author} {\bibfnamefont {V.~G.}\ \bibnamefont {Sathe}},\ }\bibfield
  {title} {\bibinfo {title} {Spin-phonon coupling in ordered double perovskites
  a$_{2}$comno$_{6}$ (a= la, pr, nd) probed by micro-raman spectroscopy},\
  }\href {https://doi.org/10.1016/j.ssc.2014.06.017} {\bibfield  {journal}
  {\bibinfo  {journal} {Solid State Commun.}\ }\textbf {\bibinfo {volume}
  {194}},\ \bibinfo {pages} {59} (\bibinfo {year} {2014})}\BibitemShut
  {NoStop}%
\bibitem [{\citenamefont {Silva}\ \emph {et~al.}(2019)\citenamefont {Silva},
  \citenamefont {Almeida}, \citenamefont {Moreira}, \citenamefont {Paniago},
  \citenamefont {Rezende},\ and\ \citenamefont
  {Paschoal}}]{silva2019vibrational}%
  \BibitemOpen
  \bibfield  {author} {\bibinfo {author} {\bibfnamefont {R.}~\bibnamefont
  {Silva}}, \bibinfo {author} {\bibfnamefont {R.}~\bibnamefont {Almeida}},
  \bibinfo {author} {\bibfnamefont {R.}~\bibnamefont {Moreira}}, \bibinfo
  {author} {\bibfnamefont {R.}~\bibnamefont {Paniago}}, \bibinfo {author}
  {\bibfnamefont {M.}~\bibnamefont {Rezende}},\ and\ \bibinfo {author}
  {\bibfnamefont {C.}~\bibnamefont {Paschoal}},\ }\bibfield  {title} {\bibinfo
  {title} {Vibrational properties and infrared dielectric features of
  gd$_{2}$comno$_{6}$ and y$_{2}$comno$_{6}$ double perovskites},\ }\href
  {https://doi.org/10.1016/j.ceramint.2018.11.168} {\bibfield  {journal}
  {\bibinfo  {journal} {Ceram. Int.}\ }\textbf {\bibinfo {volume} {45}},\
  \bibinfo {pages} {4756} (\bibinfo {year} {2019})}\BibitemShut {NoStop}%
\bibitem [{\citenamefont {Das}\ \emph {et~al.}(2019)\citenamefont {Das},
  \citenamefont {Lekshmi}, \citenamefont {Das},\ and\ \citenamefont
  {Santhosh}}]{das2019competing}%
  \BibitemOpen
  \bibfield  {author} {\bibinfo {author} {\bibfnamefont {R.~R.}\ \bibnamefont
  {Das}}, \bibinfo {author} {\bibfnamefont {P.~N.}\ \bibnamefont {Lekshmi}},
  \bibinfo {author} {\bibfnamefont {S.}~\bibnamefont {Das}},\ and\ \bibinfo
  {author} {\bibfnamefont {P.}~\bibnamefont {Santhosh}},\ }\bibfield  {title}
  {\bibinfo {title} {Competing short-range magnetic correlations, metamagnetic
  behavior and spin-phonon coupling in nd$_{2}$comno$_{6}$ double perovskite},\
  }\href {https://doi.org/10.1016/j.jallcom.2018.09.171} {\bibfield  {journal}
  {\bibinfo  {journal} {J. Alloys Compd.}\ }\textbf {\bibinfo {volume} {773}},\
  \bibinfo {pages} {770} (\bibinfo {year} {2019})}\BibitemShut {NoStop}%
\bibitem [{\citenamefont {Anshul}\ \emph {et~al.}(2020)\citenamefont {Anshul},
  \citenamefont {Kumar},\ and\ \citenamefont {Raj}}]{anshul2020raman}%
  \BibitemOpen
  \bibfield  {author} {\bibinfo {author} {\bibfnamefont {A.}~\bibnamefont
  {Anshul}}, \bibinfo {author} {\bibfnamefont {M.}~\bibnamefont {Kumar}},\ and\
  \bibinfo {author} {\bibfnamefont {A.}~\bibnamefont {Raj}},\ }\bibfield
  {title} {\bibinfo {title} {Raman and photoluminescence spectral studies in
  double perovskite epitaxial nd$_{2}$comno$_{6}$ thin films deposited by pulse
  laser deposition},\ }\href {https://doi.org/10.1016/j.ijleo.2020.164749}
  {\bibfield  {journal} {\bibinfo  {journal} {Optik}\ }\textbf {\bibinfo
  {volume} {212}},\ \bibinfo {pages} {164749} (\bibinfo {year}
  {2020})}\BibitemShut {NoStop}%
\bibitem [{\citenamefont {Balkanski}\ \emph {et~al.}(1983)\citenamefont
  {Balkanski}, \citenamefont {Wallis},\ and\ \citenamefont
  {Haro}}]{balkanski1983anharmonic}%
  \BibitemOpen
  \bibfield  {author} {\bibinfo {author} {\bibfnamefont {M.}~\bibnamefont
  {Balkanski}}, \bibinfo {author} {\bibfnamefont {R.}~\bibnamefont {Wallis}},\
  and\ \bibinfo {author} {\bibfnamefont {E.}~\bibnamefont {Haro}},\ }\bibfield
  {title} {\bibinfo {title} {Anharmonic effects in light scattering due to
  optical phonons in silicon},\ }\href
  {https://doi.org/10.1103/PhysRevB.28.1928} {\bibfield  {journal} {\bibinfo
  {journal} {Phys. Rev. B}\ }\textbf {\bibinfo {volume} {28}},\ \bibinfo
  {pages} {1928} (\bibinfo {year} {1983})}\BibitemShut {NoStop}%
\bibitem [{\citenamefont {Granado}\ \emph {et~al.}(1999)\citenamefont
  {Granado}, \citenamefont {Garc{\'\i}a}, \citenamefont {Sanjurjo},
  \citenamefont {Rettori}, \citenamefont {Torriani}, \citenamefont {Prado},
  \citenamefont {S{\'a}nchez}, \citenamefont {Caneiro},\ and\ \citenamefont
  {Oseroff}}]{granado1999magnetic}%
  \BibitemOpen
  \bibfield  {author} {\bibinfo {author} {\bibfnamefont {E.}~\bibnamefont
  {Granado}}, \bibinfo {author} {\bibfnamefont {A.}~\bibnamefont
  {Garc{\'\i}a}}, \bibinfo {author} {\bibfnamefont {J.}~\bibnamefont
  {Sanjurjo}}, \bibinfo {author} {\bibfnamefont {C.}~\bibnamefont {Rettori}},
  \bibinfo {author} {\bibfnamefont {I.}~\bibnamefont {Torriani}}, \bibinfo
  {author} {\bibfnamefont {F.}~\bibnamefont {Prado}}, \bibinfo {author}
  {\bibfnamefont {R.}~\bibnamefont {S{\'a}nchez}}, \bibinfo {author}
  {\bibfnamefont {A.}~\bibnamefont {Caneiro}},\ and\ \bibinfo {author}
  {\bibfnamefont {S.}~\bibnamefont {Oseroff}},\ }\bibfield  {title} {\bibinfo
  {title} {Magnetic ordering effects in the raman spectra of
  {La$_{1-x}$Mn$_{1-x}$O$_{3}$}},\ }\href
  {https://doi.org/10.1103/PhysRevB.60.11879} {\bibfield  {journal} {\bibinfo
  {journal} {Phys. Rev. B}\ }\textbf {\bibinfo {volume} {60}},\ \bibinfo
  {pages} {11879} (\bibinfo {year} {1999})}\BibitemShut {NoStop}%
\bibitem [{\citenamefont {Karmakar}\ \emph {et~al.}(2008)\citenamefont
  {Karmakar}, \citenamefont {Taran}, \citenamefont {Bose}, \citenamefont
  {Chaudhuri}, \citenamefont {Sun}, \citenamefont {Huang},\ and\ \citenamefont
  {Yang}}]{karmakar2008evidence}%
  \BibitemOpen
  \bibfield  {author} {\bibinfo {author} {\bibfnamefont {S.}~\bibnamefont
  {Karmakar}}, \bibinfo {author} {\bibfnamefont {S.}~\bibnamefont {Taran}},
  \bibinfo {author} {\bibfnamefont {E.}~\bibnamefont {Bose}}, \bibinfo {author}
  {\bibfnamefont {B.}~\bibnamefont {Chaudhuri}}, \bibinfo {author}
  {\bibfnamefont {C.}~\bibnamefont {Sun}}, \bibinfo {author} {\bibfnamefont
  {C.}~\bibnamefont {Huang}},\ and\ \bibinfo {author} {\bibfnamefont
  {H.}~\bibnamefont {Yang}},\ }\bibfield  {title} {\bibinfo {title} {Evidence
  of intrinsic exchange bias and its origin in spin-glass-like disordered
  l$_{0.5}$sr$_{0.5}$ mno$_{3}$ manganites (l= y, y$_{0.5}$ sm$_{0.5}$, and
  y$_{0.5}$ la$_{0.5}$)},\ }\href {https://doi.org/10.1103/PhysRevB.77.144409}
  {\bibfield  {journal} {\bibinfo  {journal} {Phys. Rev. B}\ }\textbf {\bibinfo
  {volume} {77}},\ \bibinfo {pages} {144409} (\bibinfo {year}
  {2008})}\BibitemShut {NoStop}%
\bibitem [{\citenamefont {Paccard}\ \emph {et~al.}(1966)\citenamefont
  {Paccard}, \citenamefont {Schlenker}, \citenamefont {Massenet}, \citenamefont
  {Montmory},\ and\ \citenamefont {Yelon}}]{paccard1966new}%
  \BibitemOpen
  \bibfield  {author} {\bibinfo {author} {\bibfnamefont {D.}~\bibnamefont
  {Paccard}}, \bibinfo {author} {\bibfnamefont {C.}~\bibnamefont {Schlenker}},
  \bibinfo {author} {\bibfnamefont {O.}~\bibnamefont {Massenet}}, \bibinfo
  {author} {\bibfnamefont {R.}~\bibnamefont {Montmory}},\ and\ \bibinfo
  {author} {\bibfnamefont {A.}~\bibnamefont {Yelon}},\ }\bibfield  {title}
  {\bibinfo {title} {A new property of ferromagnetic-antiferromagnetic
  coupling},\ }\href {https://doi.org/10.1002/pssb.19660160131} {\bibfield
  {journal} {\bibinfo  {journal} {Phys. Status Solidi B}\ }\textbf {\bibinfo
  {volume} {16}},\ \bibinfo {pages} {301} (\bibinfo {year} {1966})}\BibitemShut
  {NoStop}%
\bibitem [{\citenamefont {Stamps}(2000)}]{stamps2000mechanisms}%
  \BibitemOpen
  \bibfield  {author} {\bibinfo {author} {\bibfnamefont {R.}~\bibnamefont
  {Stamps}},\ }\bibfield  {title} {\bibinfo {title} {Mechanisms for exchange
  bias},\ }\href {https://doi.org/10.1088/0022-3727/33/23/201} {\bibfield
  {journal} {\bibinfo  {journal} {J. Phys. D: Appl. Phys.}\ }\textbf {\bibinfo
  {volume} {33}},\ \bibinfo {pages} {R247} (\bibinfo {year}
  {2000})}\BibitemShut {NoStop}%
\bibitem [{\citenamefont {Binek}(2004)}]{binek2004training}%
  \BibitemOpen
  \bibfield  {author} {\bibinfo {author} {\bibfnamefont {C.}~\bibnamefont
  {Binek}},\ }\bibfield  {title} {\bibinfo {title} {Training of the
  exchange-bias effect: A simple analytic approach},\ }\href
  {https://doi.org/10.1103/PhysRevB.70.014421} {\bibfield  {journal} {\bibinfo
  {journal} {Phys. Rev. B}\ }\textbf {\bibinfo {volume} {70}},\ \bibinfo
  {pages} {014421} (\bibinfo {year} {2004})}\BibitemShut {NoStop}%
\bibitem [{\citenamefont {Mishra}\ \emph {et~al.}(2009)\citenamefont {Mishra},
  \citenamefont {Radu}, \citenamefont {D{\"u}rr},\ and\ \citenamefont
  {Eberhardt}}]{mishra2009training}%
  \BibitemOpen
  \bibfield  {author} {\bibinfo {author} {\bibfnamefont {S.}~\bibnamefont
  {Mishra}}, \bibinfo {author} {\bibfnamefont {F.}~\bibnamefont {Radu}},
  \bibinfo {author} {\bibfnamefont {H.}~\bibnamefont {D{\"u}rr}},\ and\
  \bibinfo {author} {\bibfnamefont {W.}~\bibnamefont {Eberhardt}},\ }\bibfield
  {title} {\bibinfo {title} {Training-induced positive exchange bias in
  nife$/$irmn bilayers},\ }\href
  {https://doi.org/10.1103/PhysRevLett.102.177208} {\bibfield  {journal}
  {\bibinfo  {journal} {Phys. Rev. Lett.}\ }\textbf {\bibinfo {volume} {102}},\
  \bibinfo {pages} {177208} (\bibinfo {year} {2009})}\BibitemShut {NoStop}%
\bibitem [{\citenamefont {Ventura}\ \emph {et~al.}(2008)\citenamefont
  {Ventura}, \citenamefont {Araujo}, \citenamefont {Sousa}, \citenamefont
  {Veloso},\ and\ \citenamefont {Freitas}}]{ventura2008training}%
  \BibitemOpen
  \bibfield  {author} {\bibinfo {author} {\bibfnamefont {J.}~\bibnamefont
  {Ventura}}, \bibinfo {author} {\bibfnamefont {J.}~\bibnamefont {Araujo}},
  \bibinfo {author} {\bibfnamefont {J.}~\bibnamefont {Sousa}}, \bibinfo
  {author} {\bibfnamefont {A.}~\bibnamefont {Veloso}},\ and\ \bibinfo {author}
  {\bibfnamefont {P.}~\bibnamefont {Freitas}},\ }\bibfield  {title} {\bibinfo
  {title} {Training effect in specular spin valves},\ }\href
  {https://doi.org/10.1103/PhysRevB.77.184404} {\bibfield  {journal} {\bibinfo
  {journal} {Phys. Rev. B}\ }\textbf {\bibinfo {volume} {77}},\ \bibinfo
  {pages} {184404} (\bibinfo {year} {2008})}\BibitemShut {NoStop}%
\bibitem [{\citenamefont {Huang}\ \emph {et~al.}(2008)\citenamefont {Huang},
  \citenamefont {Ding}, \citenamefont {Zhang}, \citenamefont {Hou},
  \citenamefont {Yao},\ and\ \citenamefont {Li}}]{huang2008size}%
  \BibitemOpen
  \bibfield  {author} {\bibinfo {author} {\bibfnamefont {X.}~\bibnamefont
  {Huang}}, \bibinfo {author} {\bibfnamefont {J.}~\bibnamefont {Ding}},
  \bibinfo {author} {\bibfnamefont {G.}~\bibnamefont {Zhang}}, \bibinfo
  {author} {\bibfnamefont {Y.}~\bibnamefont {Hou}}, \bibinfo {author}
  {\bibfnamefont {Y.}~\bibnamefont {Yao}},\ and\ \bibinfo {author}
  {\bibfnamefont {X.}~\bibnamefont {Li}},\ }\bibfield  {title} {\bibinfo
  {title} {Size-dependent exchange bias in la$_{0.25}$ca$_{0.75}$mno$_{3}$
  nanoparticles},\ }\href {https://doi.org/10.1103/PhysRevB.78.224408}
  {\bibfield  {journal} {\bibinfo  {journal} {Phys. Rev. B}\ }\textbf {\bibinfo
  {volume} {78}},\ \bibinfo {pages} {224408} (\bibinfo {year}
  {2008})}\BibitemShut {NoStop}%
\bibitem [{\citenamefont {Moutis}\ \emph {et~al.}(2001)\citenamefont {Moutis},
  \citenamefont {Christides}, \citenamefont {Panagiotopoulos},\ and\
  \citenamefont {Niarchos}}]{moutis2001exchange}%
  \BibitemOpen
  \bibfield  {author} {\bibinfo {author} {\bibfnamefont {N.}~\bibnamefont
  {Moutis}}, \bibinfo {author} {\bibfnamefont {C.}~\bibnamefont {Christides}},
  \bibinfo {author} {\bibfnamefont {I.}~\bibnamefont {Panagiotopoulos}},\ and\
  \bibinfo {author} {\bibfnamefont {D.}~\bibnamefont {Niarchos}},\ }\bibfield
  {title} {\bibinfo {title} {Exchange-coupling properties of
  la$_{1-x}$ca$_{x}$mno$_{3}$ ferromagnetic$/$antiferromagnetic multilayers},\
  }\href {https://doi.org/10.1103/PhysRevB.64.094429} {\bibfield  {journal}
  {\bibinfo  {journal} {Phys. Rev. B}\ }\textbf {\bibinfo {volume} {64}},\
  \bibinfo {pages} {094429} (\bibinfo {year} {2001})}\BibitemShut {NoStop}%
\bibitem [{\citenamefont {Niebieskikwiat}\ and\ \citenamefont
  {Salamon}(2005)}]{niebieskikwiat2005intrinsic}%
  \BibitemOpen
  \bibfield  {author} {\bibinfo {author} {\bibfnamefont {D.}~\bibnamefont
  {Niebieskikwiat}}\ and\ \bibinfo {author} {\bibfnamefont {M.}~\bibnamefont
  {Salamon}},\ }\bibfield  {title} {\bibinfo {title} {Intrinsic interface
  exchange coupling of ferromagnetic nanodomains in a charge ordered
  manganite},\ }\href {https://doi.org/10.1103/PhysRevB.72.174422} {\bibfield
  {journal} {\bibinfo  {journal} {Phys. Rev. B}\ }\textbf {\bibinfo {volume}
  {72}},\ \bibinfo {pages} {174422} (\bibinfo {year} {2005})}\BibitemShut
  {NoStop}%
\bibitem [{\citenamefont {Giri}\ \emph {et~al.}(2011)\citenamefont {Giri},
  \citenamefont {Patra},\ and\ \citenamefont {Majumdar}}]{giri2011exchange}%
  \BibitemOpen
  \bibfield  {author} {\bibinfo {author} {\bibfnamefont {S.}~\bibnamefont
  {Giri}}, \bibinfo {author} {\bibfnamefont {M.}~\bibnamefont {Patra}},\ and\
  \bibinfo {author} {\bibfnamefont {S.}~\bibnamefont {Majumdar}},\ }\bibfield
  {title} {\bibinfo {title} {Exchange bias effect in alloys and compounds},\
  }\href {https://doi.org/10.1088/0953-8984/23/7/073201} {\bibfield  {journal}
  {\bibinfo  {journal} {J. Phys.: Condens. Matter}\ }\textbf {\bibinfo {volume}
  {23}},\ \bibinfo {pages} {073201} (\bibinfo {year} {2011})}\BibitemShut
  {NoStop}%
\end{thebibliography}%

\end{document}